\documentstyle[12pt]{article}
\begin{document}
%\draft

% below commands are for enumerating eqs according to sections
%\newcounter{eq}[section]
%\newcommand{\set}{\stepcounter{eq}
%\renewcommand{\theequation}{\mbox{\arabic{section}.\arabic{eq}}}}

\hsize=6.15in
\vsize=8.2in
\hoffset=-0.42in
\voffset=-0.3435in

\normalbaselineskip=24pt\normalbaselines

\noindent {\it Journal of Computational Neuroscience}, in press.

\vspace{0.6cm}

\begin{center}
{\large \bf Systems level circuit model of {\it C. elegans} undulatory 
locomotion: mathematical modeling and molecular genetics}
\end{center}

\vspace{0.3cm}

\begin{center}
{Jan Karbowski$^{1,2,*}$, Gary Schindelman$^{1}$,
Christopher J. Cronin$^{1}$, Adeline Seah$^{1}$, \\
and Paul W. Sternberg$^{1}$ }
\end{center}

\vspace{0.12cm}

\begin{center}
$^{1}$ {\it Howard Hughes Medical Institute and
Division of Biology 156-29}, \\
$^{2}$ {\it  Sloan-Swartz Center for Theoretical Neurobiology,
Division of Biology 216-76, \\
California Institute of Technology,
Pasadena, CA 91125, USA } 
\end{center}

%\date{\today}

\vspace{2.65cm}

\noindent {\bf Abbreviated title:} Circuit model of  
{\it C. elegans} locomotion.

\vspace{0.35cm}

\noindent {\bf Keywords}: C. elegans, circuit model for movement, 
oscillations, mechanosensory feedback, GABA, acetylcholine, calcium, 
myosin, mutants.

\vspace{0.4cm}

\noindent $^{*}$ Corresponding author at: jkarb@its.caltech.edu.
Phone: (626)-395-5840.

\newpage

%\widetext
\begin{abstract}
To establish the relationship between locomotory behavior and dynamics of 
neural circuits in the nematode {\it C. elegans} we combined molecular and 
theoretical approaches. In particular, we quantitatively analyzed the motion 
of {\it C. elegans} with defective synaptic GABA and acetylcholine 
transmission, defective muscle calcium signaling, and defective muscles 
and cuticle structures, and compared the data with our systems level circuit
model. The major experimental findings are: 
(i) anterior-to-posterior gradients of body bending flex for almost
all strains both for forward and backward motion, and for neuronal 
mutants, also analogous weak gradients of undulatory frequency,
(ii) existence of some form of neuromuscular (stretch receptor) feedback,
(iii) invariance of neuromuscular wavelength,  
(iv) biphasic dependence of frequency on synaptic signaling, and
(v) decrease of frequency with increase of the muscle time 
constant. Based on (i) we hypothesize that the Central Pattern Generator 
(CPG) is located in the head both for forward and backward motion. 
Points (i) and (ii) are the starting assumptions for our theoretical model, 
whose dynamical patterns are qualitatively insensitive to the details of 
the CPG design if stretch receptor feedback is sufficiently strong 
and slow. The model reveals that stretch receptor coupling in the body 
wall is critical for generation of the neuromuscular wave.
Our model agrees with our behavioral data (iii), (iv), and (v), and with 
other pertinent published data, e.g., that frequency is an increasing 
function of muscle gap-junction coupling. 

\end{abstract}

%\narrowtext 
%\maketitle 

%\begin{narrowtext}

%\maketitle

\newpage

\noindent {\bf\Large Introduction}

\vspace{0.6cm}

{\it Caenorhabditis elegans} nematode worms, with a small nervous system
comprising only 302 neurons (White et al, 1986), move by generating an 
oscillatory neuromuscular wave that alternates dorsal and ventral muscles 
(Brenner, 1974; Chalfie et al, 1985; Karbowski et al, 2006). 
The molecular, cellular, and 
network mechanisms of this oscillatory spatio-temporal activity are 
virtually unknown. Their understanding may provide insight about the 
relationship between neuromuscular dynamics and how behavior is created 
in these extensively genetically studied animals 
(Bargmann, 1998; Hobert, 2003; de Bono and Maricq, 2005; Gray et al, 2005), 
and might be potentially relevant for other locomotory systems. 
It has proven difficult to address these questions using standard
electrophysiological techniques in {\it C. elegans} because of its small 
neural sizes (Francis et al, 2003). 
In this paper, we use a combination of genetic perturbations, 
behavioral assays combined with a quantitative tracking system, 
and mathematical modeling to decipher dynamical properties of a 
detailed neuromuscular circuit relevant for {\it C. elegans} movement.
By combining these approaches and building a mathematical
circuit model we seek to bridge the gap between molecular/cellular
and systems level understandings of undulatory locomotion (Fig. 1).

We investigate specific questions related to the mechanisms that control 
body undulations and coordination:
(i) How do different elements in the {\it C. elegans} neuromuscular
circuit interact to produce both body oscillations and neuromuscular
wave? In particular, where is the primary oscillatory signal generated?
(ii) How does mechanosensory feedback (stretch receptor coupling) affect
locomotion? Does it play any role in generating a neuromuscular wave?
If so, how does its strength affect the wavelength (intermuscle phase
lag) of muscle contractions? (iii) How does synaptic coupling between
neurons affect locomotory rhythm? Is there any qualitative difference
between excitation and inhibition on oscillatory frequency? Are there
optimal values for the synaptic couplings? (iv) How does gap-junction
coupling between body-wall muscles and structural defects in muscles
and cuticle affect movement?

To address these questions we employed a parallel approach of collecting
experimental data and making model predictions in an iterative manner.
In particular, we quantitatively analyzed the motion of different neuronal 
and non-neuronal mutants by measuring their kinematic parameters (Fig. 1B) 
and relating them to the dynamic properties of our circuit model. 
Investigated mutants included worms with decreased and increased synaptic 
GABA transmission, altered acetylcholine transmission, altered levels 
of calcium signaling in muscles, and worms with structural defects in 
muscles (myosin) and cuticle.

%\vspace{1.5cm}

\newpage

\noindent{\bf\Large Materials and Methods}

\vspace{0.8cm}

\noindent{\bf Experimental part}

\vspace{0.4cm}

\noindent {\it Description of analyzed mutants.}
We examined the locomotion of several 
{\it C. elegans} mutants with affected inhibitory synaptic transmission
(GABA), excitatory synaptic transmission (acetylcholine), and muscular 
function (calcium channels and myosin). We investigated
mutants with both decreased and increased GABA function. For mutations
effectively decreasing inhibition we studied: {\it unc-25(e156)}, which
encodes glutamic acid decarboxylase (GAD) - the biosynthetic enzyme for GABA
production (Jin et al, 1999); {\it unc-46(e177)} ({\it him-5(e1490)} was 
in the background), which presumably plays a modulatory role in GABA packaging 
into vesicles (Schuske et al, 2004); and {\it unc-18(e81)} mutants carrying 
the {\it syEx995[cho-1::unc-18::yfp]} transgene. {\it unc-18} functions as a 
facilitator of vesicles docking at presynaptic neurons (Weimer et al, 2003) 
and is expressed in both cholinergic and GABAergic neurons 
(Gengyo-Ando et al, 1993). Therefore, in the {\it unc-18(e81)}
strain carrying {\it syEx995[cho-1::unc-18::yfp]} we have restored 
{\it unc-18} function specifically to cholinergic neurons using the 
{\it cho-1} promoter (Okuda et al, 2000), leaving the GABAergic neurons 
non-functional. For a mutation effectively increasing inhibition we studied 
{\it slo-1(js118)} mutants with extrachromosomal {\it Pacr-2::slo-1(+)}
transgene. SLO-1 encodes a Ca$^{2+}$ activated K$^{+}$ channel that 
presumably counteracts quantal synaptic neurotransmission both in excitatory
and inhibitory neurons (Wang et al, 2001), and removal of this channel
(in {\it slo-1} mutants) leads to an increased neurotransmission above
wild-type level. {\it slo-1(js118)} worms with the {\it Pacr-2::slo-1(+)} 
transgenes (syEx996, syEx988, and syEx991) have elevated neurotransmitter
release in inhibitory neurons, but they have restored wild-type 
release levels in excitatory neurons because 
{\it Pacr-2::slo-1(+)} drives expression of wild-type {\it slo-1} in
these neurons (Davies et al, 2003). We also studied three types of
strains with presumably elevated acetylcholine (excitatory) synaptic
signaling and wild-type level of GABA: {\it slo-1(js118); Punc-25::slo-1},
$\;\;$ {\it slo-1(js118); Punc-17::slo-1}, $\;\;$ 
and {\it slo-1(js118); Pcho-1::slo-1}.

For mutations affecting muscular calcium function, we studied {\it unc-68} 
mutants, as well as {\it egl-19(n582)} rescued with an
{\it unc-119::egl-19::yfp} transgene.
{\it unc-68} gene encodes ryanodine receptor channels that gate the 
release of calcium ions from internal stores in both body-wall muscle 
cells and in pharyngeal muscle (Maryon et al, 1996; Maryon et al, 1998). 
We analyzed the locomotion of loss-of-function {\it unc-68(r1158)} 
worms, which exhibit incomplete flaccid paralysis. The {\it egl-19} gene 
plays an important role in regulating muscle excitation and encodes the 
$\alpha 1$ subunit of a homologue of vertebrate L-type voltage-activated 
Ca$^{2+}$ channels (Lee et al, 1997). Similarly, we analyzed the locomotion 
of  {\it egl(n582)} mutants carrying an {\it unc-119::egl-19::yfp} 
transgene, which restores {\it egl-19} function specifically in neurons 
using the {\it unc-119} promoter to drive pan-neuronal expression of 
{\it egl-19} (Maduro and Pilgrim, 1995). This partial loss-of-function
strain have slow muscle depolarization and feeble contractions 
(Lee et al, 1997; Jospin et al, 2002).

For mutations affecting muscle and cuticle structure we analyzed 
{\it unc-54}, {\it sqt-1}, and BE109 worms.
{\it unc-54} worms carry mutations in the head region of 
body-wall muscle myosin. We analyzed the following strains: 
RW130 {\it unc-54(st130)}, RW132 {\it unc-54(st132)}, 
RW134 {\it unc-54(st134)}, RW135 {\it unc-54(st135)}, 
RW5008 {\it unc-54(s95)}, and BC347 {\it unc-54(s74)}. 
These mutations are hypothesized to alter the contraction-relaxation 
cycle of the myosin-actin crossbridge formation by increasing its 
duration (Moerman and Fire, 1997). {\it sqt-1} gene encodes cuticle 
collagen and we analyzed two strains: BE101 {\it sqt-1(sc101)} and BE103 
{\it sqt-1(sc103)}. BE109 mutants have defective cuticle struts 
(J. Kramer, pers. communication).

Construction of plasmids and strains is described in the Supplementary 
Information.

\vspace{1cm}

\noindent {\it Locomotory data extraction and processing.}
We video recorded and digitized the motion of young adult hermaphrodite 
worms (mutants and wild-type) 15-20 hr post mid-L4 developmental stage on 
an agar plate with thin film of {\it E. coli} OP50 bacteria and LB media. 
The video recording and data extraction was done using a device 
specially designed for studying {\it Caenorhabditis\/} locomotion 
(Cronin et al, 2005). We collected 5 minutes of video per worm, extracting
digital locomotion data from the middle 4 minutes.
Such 4 minute windows average over many possible sensory influences that 
can vary among worms and thus statistically minimize the variability 
of external conditions. From these data we derived distributions of the
undulatory frequency and bending flex by dividing worm's body into 12
sections which served as worm's system of coordinates. Local flex
angle was determined as described in Fig. 1B.
To derive the bending frequencies for each articulation point we first
calculate a time-ordered matrix of angles between each pair of adjacent body
sections (that is, at each articulation point).  Next we apply Matlab's
spectrogram function (specgram) to the vector of changing bend angles at
each articulation point, which yields a time-ordered matrix of the relative
energies of the changing bend angles' component frequencies.  The highest
magnitude, non-constant component frequency from each time window is that
window's characteristic bending frequency. Frequencies and flexes for
each articulation point in Figs. 3-6 are population averages. In Figs. 5
and 6 we used non-parametric sign test for paired samples to
determine if average frequencies between different mutants are
significantly different.

%\newpage
\vspace{1cm}

\noindent{\bf Theoretical part}

\vspace{0.4cm}

\noindent {\it Circuit model.}
All known locomotory systems use some form of a Central Pattern Generator
(CPG) circuit to generate rhythm (Grillner, 1975; Delcomyn, 1980; Marder and
Calabrese, 1996; Nusbaum and Beenhakker, 2002). Mechanosensory feedback
usually alters the rhythm; however, its importance varies in different
animals and even in the same animal under different conditions 
(Marder et al, 2005). At one extreme, it can play a minor modulatory role
(Grillner, 1975; Delcomyn, 1980), at the other, it can be critical for
body coordination (Friesen and Cang, 2001; Akay et al, 2004).

The global circuit responsible for {\it C. elegans} forward locomotion 
is constructed based on neuroanatomical data 
(Chalfie et al, 1985; White et al, 1986). 
It is composed of the head and body wall neural networks (Fig. 2A,B) 
connected partly via two interneurons AVB and PVC, which are implicated 
as controllers of forward movement (Chalfie et al, 1985), and partly 
via muscle gap junctions.  AVB and PVC receive massive synaptic input from 
the head neurons and target only excitatory motor neurons (type B neurons: 
ventral VB and dorsal DB) in the body wall (White et al, 1986).
AVB connects them via gap junctions and PVC via chemical synapses. We assume,
following theoretical predictions (Wicks et al, 1996), that the latter 
synapses are excitatory.

The real head neural network is composed of about 200 neurons with
mostly unknown polarities and complicated wiring diagram (White el al, 1986).
Inclusion of all of these elements in our circuit model would increase
its complexity (the number of parameters) by at least an order of magnitude,
which would obscure the basic mechanisms. Instead, we decided to depict
the head network in a simplified manner with lesser but essential number 
of elements. This simplified approach has an advantage of being conceptually
comprehensible. Thus our head neural network is comprised of head 
interneurons and ventral and dorsal head motor neurons (Fig. 2B). 
The head motor neurons contain both excitatory neurons (White et al, 1986) 
and a pair of GABA-ergic inhibitory RME neurons (Schuske et al, 2004). 
Because connectivity patterns between actual head interneurons form
numerous feedback loops (e.g., AIZ $\rightarrow$ AIA $\leftrightarrow$
AWA $\rightarrow$ AIZ and  RIB $\leftrightarrow$ RIG $\leftrightarrow$
URY $\rightarrow$ RIB; see White et al, 1986), we mimicked this phenomenon 
by two (dorsal and ventral) interneuron loops 
X$\rightarrow$ Y$\rightarrow$ Z$\rightarrow$ X.
These loops can in principle generate a primary rhythm in the head, 
which is then imposed on downstream networks along the body via muscle 
coupling and/or AVB and PVC interneurons. The loop X, Y, Z is the minimal
circuit in the {\it C. elegans} head interneurons that can generate 
robust oscillations. We include mechanosensory 
feedback in our head network
model following suggestions that a pair of head interneurons SAA may
contain stretch receptors on their long dendrites (White et al, 1986).

The body wall forward motion network (White et al, 1986) is composed of 
ventral and dorsal sub-circuits that contain motor neurons, body wall 
longitudinal muscles, and feedback loops associated with mechanosensation 
(Fig. 2A). The neuronal part of each sub-circuit consists of several 
cholinergic excitatory B motor neurons (ventral VB and dorsal DB), 
and several GABA-ergic inhibitory motor neurons (type D neurons: ventral
VD and dorsal DD) (Chalfie et al, 1985; McIntire et al, 1993;
Schuske et al, 2004). Apart from gap junctions from AVB and chemical
synapses from PVC, the excitatory B neurons receive synaptic input from 
DVA interneuron (DVA synapses on some B neurons, see White et al, 1986).
Anatomically, the role of this interneuron in locomotion is not clear,
and it was suggested (Li et al, 2006) that it might serve as a neuromodulator
involved in detecting body stretch, with no definite synaptic polarity.  
We assume that the overall influence of this neuron is inhibitory. 
This inhibitory input is necessary 
to prevent spontaneous oscillations in the body-wall circuit when
it is disconnected from the head.
Coupling between ventral and dorsal motor sub-circuits is mediated 
partly via excitatory synapses terminating on inhibitory D motor neurons of 
the complementary circuit (Fig. 2A), and partly via AVB gap-junctions.
An effective cross-inhibition 
(B $\rightarrow$ D) between the two sub-circuits
amplifies their alternating oscillatory activities (Fig. 2A, and below). 
Neighboring motor neurons, as well as neighboring muscles are connected
via gap junctions (D. Hall, pers. commun). Mechanosensory feedback in the 
body wall, similarly as in the head, is associated 
with hypothetical stretch receptors. Following the Byerly and Russell 
hypothesis (in Chalfie and White, 1988), 
we assume that stretch receptors are located on the extended 
dendritic processes of the excitatory B motor neurons (VB and DB). 
In the forward motion circuit, all these dendrites are 
directed posteriorly. We assume that mechanosensors transduce local 
body stretch along excitatory motor neuron dendrites 
to the somas of these neurons by polarizing them, since mechano-stimulation
is known to increase cell excitability (Goodman and Schwarz, 2003).
Thus, mechanosensory feedback coupling along the body wall is non-local, 
excitatory, and unidirectional in nature.

The theoretical part of this study deals almost exclusively with 
forward motion, which is a dominant behavior in {\it C. elegans}.
Backward motion is considered only briefly in the context of neuromuscular
wave directionality (see below). The backward motion is controlled by
a complementary system of 3 interneurons (AVA, AVD, AVE) and type A
excitatory motor neurons (ventral VA and dorsal DA). The latter neurons have
long dendrites directed anteriorly, i.e., opposite to the corresponding
motor neurons in the forward motion circuit (Chalfie et al, 1985).
The behavioral transitions between forward and backward motion will be
addressed in a future study.

\vspace{1cm}

\noindent {\it Equations used in the theoretical model.}
We assume that the activities of motor neurons are graded, since
experimental data show the absence of action potentials in the sensory 
neuron ASER (Goodman et al, 1998), which is consistent with the lack of
voltage-activated sodium channels in the {\it C. elegans} genome 
(Bargmann, 1998). Additional support for this assumption comes from 
neurophysiological recordings in a related, larger, nematode {\it Ascaris}, 
which has a similar nervous system, and lacks action potentials in all 
recorded neurons (Davis and Stretton, 1989). Below we provide equations 
governing activities of all neurons, muscles, and stretch receptors
included in the model. These equations were solved on a Linux-system 
computer using a standard second-order Runge-Kutta method in Fortran 
(F77).

The dynamics of head interneurons controlling the ventral side of the head 
are represented by:

\begin{eqnarray}
\tau_{x}\frac{d X_{v}}{dt} = - X_{v} + c_{1} +  w_{xs}{\bf H_{xs}}(S_{hv})
 - w_{xz}{\bf H_{xz}}(Z_{v})   \nonumber \\
- w_{xi}{\bf H_{xi}}(I_{hv}) + g_{avb}(V_{avb}-X_{v}),
\end{eqnarray}
\begin{eqnarray}
\tau_{y}\frac{d Y_{v}}{dt} = - Y_{v} + w_{yx}{\bf H_{yx}}(X_{v})
+ g_{avb}(V_{avb}-Y_{v}),
\end{eqnarray}
\begin{eqnarray}
\tau_{z}\frac{d Z_{v}}{dt} = - Z_{v} + w_{zy}{\bf H_{zy}}(Y_{v})
 + w_{zx}{\bf H_{zx}}(X_{d}) + g_{avb}(V_{avb}-Z_{v}),
\end{eqnarray}\\
where $X_{v}$ and $Y_{v}$ denote the activities of two excitatory 
ventral head interneurons, $X_{d}$ denotes the activity of a corresponding
head dorsal interneuron, 
$Z_{v}$ is the activity of ventral inhibitory interneuron, and
$\tau_{x}$, $\tau_{y}$, $\tau_{z}$ are corresponding time constants. 
$S_{hv}$ is the head stretch receptor activity situated on the ventral 
interneuron $X_{v}$, and $I_{hv}$ is the membrane potential of head
inhibitory ventral motor neuron.
$V_{avb}$ is the membrane potential of AVB interneuron and $g_{avb}$ is
the gap junction strength between AVB and $X$, $Y$, and $Z$. 
The parameter $c_{1}$ denotes
some input coming from other head neurons and it can be either excitatory
(positive) or inhibitory (negative), depending on the worm's ``internal
state''. The parameters $w_{\alpha\beta}$ characterize the strength of 
chemical synapses between circuit elements from $\beta$ to $\alpha$, 
and most of the genetic mutations we examined affect precisely these 
parameters. The functions ${\bf H_{\alpha\beta}}$ are positively
defined sigmoidal functions of the form: ${\bf H_{\alpha\beta}}(x)= 
\left(1 + \tanh[(x-\theta_{\alpha\beta})/\eta_{\alpha\beta}]\right)$, 
where $\theta_{\alpha\beta}$ is the threshold for activation
and $\eta_{\alpha\beta}$ characterizes the steepness of the nonlinearity. 
The equations for the dorsal side of the head are analogous
with a subscript substitution $v \leftrightarrow d$.

The dynamics of the ventral side head motor neurons,
muscles, and stretch receptor feedback are given by:

\begin{eqnarray}
\tau_{e}\frac{d E_{hv}}{dt} = - E_{hv} + w_{ey}{\bf H_{ey}}(Y_{v}), 
\end{eqnarray}
\begin{eqnarray}
\tau_{i}\frac{d I_{hv}}{dt} = - I_{hv} + w_{ie}{\bf H_{ie}}(E_{hv}), 
\end{eqnarray}
\begin{eqnarray}
\tau_{m}\frac{d M_{hv}}{dt} = - M_{hv} + w_{mm}{\bf H_{mm}}(M_{hv}) 
 + {\tilde w}_{me}{\bf H_{me}}(E_{hv})    \nonumber  \\
 - {\tilde w}_{mi}{\bf H_{mi}}(I_{hd}) + g_{m}(M_{v,1}-M_{hv}),  
\end{eqnarray}
\begin{eqnarray}
\tau_{s}\frac{d S_{hv}}{dt} = - S_{hv} + w_{sm}
\left[{\bf H_{sm}}(M_{hd}-M_{hv}) - 1\right],
\end{eqnarray}\\
where $E_{hv}$ and $I_{hd}$ are the membrane potentials of head ventral 
excitatory and dorsal inhibitory motor neurons respectively, 
$M_{hv}$ and $M_{hd}$ are head ventral and dorsal muscle membrane potentials, 
and $\tau_{e}$, $\tau_{i}$,
$\tau_{m}$, $\tau_{s}$ are their respective time constants. 
The steady-state value of the stretch receptor activity $S_{hv}$ should be 
proportional to the degree of head bending, which we assume is a sigmoidal
function of the differences in ventral and dorsal muscle activities. 
Note that stretch receptors are activated if there is initial asymmetry
in activities of ventral and dorsal muscles.
Apart from synaptic input, a muscle cell is also activated by 
a voltage-activated calcium channels, which in the model is represented 
by a self-promoting ${\bf H_{mm}}(M_{hv})$ term with the coupling 
$w_{mm}$ characterizing the calcium signaling strength. Identical
equations describing the dorsal side are obtained with a subscript 
substitution $hv \leftrightarrow hd$.

The dynamics of the two forward motion interneurons AVB and PVC are given
by:

\begin{eqnarray}
\tau_{avb}\frac{d V_{avb}}{dt} = - V_{avb} + w_{avb,x}
\left[{\bf H_{avb,x}}(X_{v}+Y_{v}) + {\bf H_{avb,x}}(X_{d}+Y_{d})\right]
\nonumber \\
+ w_{avb,pvc}{\bf H_{avb,pvc}}(V_{pvc})    
+ g_{avb,e}\sum_{i=1}^{N}\left(E_{v,i} + E_{d,i} - 2V_{avb}\right)
\nonumber \\  
+ g_{avb}\left(X_{v}+Y_{v}+Z_{v} + X_{d}+Y_{d}+Z_{d} - 6V_{avb}\right), 
\end{eqnarray}
\begin{eqnarray}
\tau_{pvc}\frac{d V_{pvc}}{dt} = - V_{pvc} 
+ w_{pvc,x}\left[{\bf H_{pvc,x}}(X_{v}+Y_{v}) + {\bf H_{pvc,x}}(X_{d}+Y_{d})
\right],
\end{eqnarray}\\
where $V_{avb}$ and $V_{pvc}$ are the membrane potentials of AVB and PVC
interneurons and $\tau_{avb}, \tau_{pvc}$ are their respective time constants.
$E_{v,i}$ and $E_{d,i}$ denote the membrane potentials of the $i$th
ventral (VB type) and dorsal (DB type) excitatory motor neurons,
respectively. The parameter $g_{avb,e}$ is the gap-junction coupling between 
AVB and the body wall excitatory motor neurons. $N$ denotes the numbers
of excitatory and inhibitory motor neurons on both sides of the body wall;
in simulations we take $N=8$.

The ventral side of the body wall nervous system is represented by a set of 
equations:

\begin{eqnarray}
\tau_{e}\frac{d E_{v,i}}{dt} = - E_{v,i} + c_{2} 
+ w_{e,pvc}{\bf H_{e,pvc}}(V_{pvc}) + w_{es}{\bf H_{es}}(S_{v,i+1})  
\nonumber \\ 
+ g_{avb,e}(V_{avb}-E_{v,i})  
+ g_{e}\left(E_{v,i+1} + E_{v,i-1} - 2E_{v,i}\right), 
\end{eqnarray}
\begin{eqnarray}
\tau_{i}\frac{d I_{v,i}}{dt} = - I_{v,i} + w_{ie}{\bf H_{ie}}(E_{d,i})
+ g_{i}\left(I_{v,i+1} + I_{v,i-1} - 2I_{v,i}\right),  
\end{eqnarray}
\begin{eqnarray}
\tau_{m}\frac{d M_{v,i}}{dt} = - M_{v,i} + w_{mm}{\bf H_{mm}}(M_{v,i}) 
+ w_{me}{\bf H_{me}}(E_{v,i}) 
\nonumber \\
- w_{mi}{\bf H_{mi}}(I_{v,i})
+ g_{m}\left(M_{v,i+1} + M_{v,i-1} - 2M_{v,i}\right),   
\end{eqnarray}
\begin{eqnarray}
\tau_{s}\frac{d S_{v,i}}{dt} = - S_{v,i}
+ w_{sm}\left[{\bf H_{sm}}(M_{d,i}-M_{v,i}) - 1\right],
\end{eqnarray}\\
where $I_{v,i}$ is the membrane potential of the $i$th ventral 
inhibitory motor neuron (VD type), and $S_{v,i}$ is the stretch receptor 
activity related, as in the head, to the relative activities of the ventral 
$M_{v,i}$ and dorsal $M_{d,i}$ muscles. 
By the nature of connectivity in our network,
$M_{v,0}\equiv M_{hv}$ and $M_{d,0}\equiv M_{hd}$.
Note that the $i$th excitatory motor neuron $E_{v,i}$ (corresponding to 
VB type) is affected by the stretch receptors at $(i+1)$th position in 
the forward motion circuit. 
(In the backward motion circuit $E_{v,i}$ corresponds to VA neuron and is
affected by the stretch receptor at $(i-1)$th position.) 
The parameters $g_{e}$, $g_{i}$, and $g_{m}$ are
the gap-junction couplings among excitatory motor neurons, inhibitory
motor neurons, and among muscles, respectively.
The parameter $c_{2}$ denotes inhibition coming to the excitatory motor 
neurons (VB, DB) from the DVA interneuron, and its negative value
prevents spontaneuos oscillations in the body wall circuit when it is
disconnected from the head circuit. These spontaneous oscillations
emerge because of the numerous closed loops in the body wall, which
contains both excitatory and inhibitory elements (VB $\rightarrow$ 
DD $\rightarrow$ $M_{d}$ $\rightarrow$ $S_{d}$ $\rightarrow$ DB $\rightarrow$ 
VD $\rightarrow$ $M_{v}$ $\rightarrow$ $S_{v}$ $\rightarrow$ VB), see Fig. 2A.
Identical equations describe a dorsal 
part of the body wall circuit, with a substitution $v \leftrightarrow d$. 
The ventral inhibitory motor neuron $I_{v,i}$ (VD type) is activated by 
the activity of the dorsal excitatory motor neuron $E_{d,i}$ (DB type).

In simulations we rescale uniformly almost all synapses both in the head and
in the body wall by two factors: $q_{ex}$ corresponding to excitatory
synapses, and $q_{in}$ corresponding to inhibitory GABA synapses. Both
parameters are variable and characterize global levels of excitation
and inhibition, respectively. The values of parameters used in simulations
are presented in Table 1.
Values of these parameters were chosen arbitrarily and a global
circuit function (i.e. sustained oscillations), to a large extent, 
does not depend on their precise values. Oscillations in the circuit
emerge for $c_{1}$ between 0.5 and 2.5.
Models A and B differ only in one aspect, i.e., the degree
of nonlinearity in interactions (in functions {\bf H}) between head 
interneurons $X, Y, Z$.
Model A was obtained by taking the corresponding steepness parameters
$\eta$ to be small (high nonlinearity), whereas model B was obtained 
by increasing values of these parameters (low nonlinearity).
The rest of the parameters are the same for both models.
In all theoretical figures we used the parameters from Table 1,
unless indicated otherwise.

The Fortran code for simulating of Eqs. (1)-(13) will be deposited
in {\it ModelDB} database.

\newpage
%\vspace{2.5cm}

\noindent {\bf\Large Results}

\vspace{1cm}

To begin an analysis of how the circuit controls locomotion, we chose 
specific genetic perturbations that affect easily identifiable classes of 
parameters in the neuroanatomical circuit. These parameters can be 
manipulated in our theoretical model, enabling a comparison between theory 
and experiment.

\vspace{0.3cm}

\noindent {\bf Experimental results}

\vspace{0.3cm}

The major experimental findings detailed below are: (i) anterior-to-posterior 
spatial gradients of flex angle for almost all strains, and for 
neuronal mutants, also analogous weak gradients of frequency, both for 
forward and backward motion, (ii) cuticle and muscle (non-neuronal) mutants 
affect locomotion characteristics, suggesting the existence of some
form of mechanosensory feedback, (iii) invariance of wavelength 
characterizing the neuromuscular wave, (iv) biphasic dependence of 
undulatory frequency on synaptic GABA signaling, and a plausible biphasic
dependence on acetylcholine signaling, and (v) strong decrease of
undulatory frequency with increasing muscle time constant characterizing
the cycle of myosin-actin crossbridge formation.

\vspace{0.5cm}

\noindent {\it Spatial distribution of undulatory frequency and bending flex.}
The spatial distribution of some locomotory parameters may provide clues about 
how the motion is generated and coordinated in {\it C. elegans} worms. 
We measured frequency of undulations and bending flex,
defined as a maximal local bending angle of worm's body,
as a function of their distance from the head. The frequency for wild-type 
and most non-neuronal mutants is either constant or nearly constant 
both for forward and backward motion (Fig. 3A,B, and Supplementary Table T1), 
with the exception of {\it sqt-1} (lacking one cuticle collagen),
for which the frequency decreases significantly toward the tail
(e.g., in {\it sqt-1(sc103)}, the head and tail oscillate at respectively 
$0.29\pm 0.07$ Hz and $0.22\pm 0.09$ Hz in forward motion, and
corresponding values in backward motion are: $0.22\pm 0.05$ Hz 
and $0.12\pm 0.04$ Hz; see Suppl. Table T1).
In contrast, all neuronal mutants display a monotonic decay of the 
frequency directed posteriorly, in some cases up to 50 $\%$ (see Suppl.
Table T1). Specifically, we obtained the following frequencies in the
head and tail for neuronal mutants. 
For {\it unc-25(e156)}: $0.36\pm 0.07$ Hz
in the head and $0.23\pm 0.06$ Hz in the tail (forward motion), and
$0.20\pm 0.08$ Hz in the head and $0.13\pm 0.03$ Hz in the tail
(backward motion);    
for {\it unc-46(e177)}: $0.29\pm 0.03$ Hz
in the head and $0.16\pm 0.07$ Hz in the tail (forward motion), and
$0.26\pm 0.02$ Hz in the head and $0.13\pm 0.04$ Hz in the tail
(backward motion);    
for {\it unc-18(e81); cho-1::unc-18:yfp}: $0.20\pm 0.08$ Hz
in the head and $0.09\pm 0.04$ Hz in the tail (forward motion), and
$0.16\pm 0.05$ Hz in the head and $0.06\pm 0.02$ Hz in the tail
(backward motion);    
for {\it slo-1(js118)}: $0.38\pm 0.07$ Hz
in the head and $0.29\pm 0.08$ Hz in the tail (forward motion), and
$0.32\pm 0.07$ Hz in the head and $0.16\pm 0.06$ Hz in the tail
(backward motion). These data indicate that for neuronal mutants 
tail oscillations have substantially smaller frequencies than head 
oscillations.

The behavior of the flex is more universal (Fig. 3C,D and Suppl. Table T1). 
For almost all mutants and wild-type worms, and for both forward and
backward motion, there is a pronounced decay of the flex from the head up 
to some point close to the tail. From that point the flex either stays 
constant or slightly increases towards the tail. Overall, the flex and
position exhibit statistically significant strong negative correlations
(Fig. 3 C,D). Thus, surprisingly, the opposite motions yield similarly 
directed anterior-to-posterior gradients of the flex.

\vspace{0.5cm}

\noindent {\it Some non-neuronal mutants dramatically alter frequency.}
We find that it is relatively easy to change the undulatory frequency 
of worms by genetically modifying some of non-neuronal parameters. 
As an example, mutants with 
affected cuticle function (collagen defective {\it sqt-1} mutants or strut 
defective BE109 mutants) and mutants with altered muscle function 
(myosin defective {\it unc-54} mutants) exhibit different locomotory outputs 
than the wild-type worms, with substantially reduced frequency (Fig. 4). 
Also, results (see below) on mutants with defects in muscle calcium 
signaling ({\it unc-68}) show that their frequency is significantly lower
than wild-type. 
The fact that non-neuronal mutants have profoundly altered locomotory
parameters, in particular frequency and flex, indicates that
there must be some feedback coupling between the nervous system of 
{\it C. elegans} and its muscular and cutico-skeletal 
systems.

\vspace{0.5cm}

\noindent {\it Conservation of wavelength.} We measured the wavelength 
of neuromuscular wave in wild type and several mutants (Table 2).
The ratio of the wavelength to the actual worm's body length is relatively
constant across different worms and appears to be identical for forward
and backward motion. Since these worms move with different frequencies,
the data in Table 2 suggest that wavelength is essentially frequency
independent.

\vspace{0.5cm}

\noindent {\it Biphasic dependence of undulatory frequency on GABA 
(inhibitory) transmission.}
We analyzed the motion of mutants defective in GABA function primarily 
in the ventral cord motor neurons. All three uncoordinated neuronal mutants - 
{\it unc-25(e156)}, {\it unc-46(e177)}, and {\it unc-18(e81)} rescued 
with transgene {\it syEx981[cho-1::unc-18::yfp]} in cholinergic neurons 
- have decreased GABA 
transmission as compared to wild type (McIntire et al, 1993; 
Weimer et al, 2003). These animals have also much lower frequencies of 
undulations than wild-type worms (Fig. 5A and Suppl. Table T1, 
$p < 0.05$).

We also studied the behavior of worms with increased GABA 
transmission. Since no known mutants have this property exclusively, 
we constructed and analyzed several transgenic strains with altered 
GABAergic and cholinergic signaling based on {\it slo-1(js118)} mutation, 
which elevates globally both types of neurotransmission by eliminating 
a particular calcium activated potassium channel (Wang et al, 2001) 
(Suppl. Table T2). By using different promoters
we constructed worms that restore either excitation or
inhibition to normal levels by making K$^{+}$ channel functional in
either cholinergic or GABA-ergic neurons.
The {\it slo-1(js118); Pacr-2::slo-1} worms have presumably rescued 
cholinergic transmission to the wild-type level, but GABA transmission
is still elevated. These worms are significantly slower ($p < 0.05$) than
wild-type worms (Fig. 5B). For consistency, we also compared 
{\it slo-1(js118); Pacr-2::slo-1} animals to two strains with presumably
wild-type levels of both GABA and acetylcholine, i.e., 
{\it slo-1(js118); Pacr-2::slo-1; Punc-25::slo-1} $\;\;$ and  
{\it slo-1(js118); Psnb-1::slo-1} (Fig. 5C). In this case
{\it slo-1(js118); Pacr-2::slo-1} are also significantly slower ($p < 0.05$).
Thus, the behavior of worms with increased GABA signaling (Figs. 5B,C)
is qualitatively similar to the behavior of worms with decreased
GABA (Fig. 5A): the undulatory frequency of worms with altered GABA
transmission, either below or above the wild-type level, is always
smaller than wild type.

To summarize, all these results clearly suggest a non-monotonic (biphasic) 
dependence of the undulatory frequency on the inhibitory 
synaptic GABA signaling in the worm.

\vspace{0.5cm}

\noindent {\it Dependence of undulatory frequency on acetylcholine
(excitatory) transmission.}
Various constructs with {\it slo-1} enable us to determine the dependence of 
frequency on excitatory coupling in {\it C. elegans} (Suppl. Table T2). 
{\it slo-1(js118); Pacr-2::slo-1} animals with presumably rescued
levels of acetylcholine transmission to the wild-type level but elevated 
GABA transmission are either slightly faster or slightly slower than the 
control {\it slo-1(js118); Pacr-2::gfp} with both increased GABA and
increased acetylcholine signaling (Fig. 6A). 
Thus, a decrease of excitatory transmission does not lead to a clear-cut 
phenotype, which might suggest a possible non-monotonic dependence of
the frequency on a global synaptic excitation. In the case of worms with 
increased acetylcholine signaling and presumably wild-type level of GABA 
(three strains: {\it slo-1(js118); Punc-25::slo-1}, $\;\;$ and
{\it slo-1(js118); Punc-17::slo-1}, $\;\;$ and 
{\it slo-1(js118); Pcho-1::slo-1}), 
we observe their slightly lower frequencies than wild type (Fig. 6B). 
This case, in turn, suggests that increasing global excitation leads 
to a slight decrease in undulatory frequency (see, however, ``Comparison
of the theory with the experimental data'' section below). 

\vspace{0.5cm}

\noindent {\it Frequency decreases sharply with increasing muscle
time constant.}
Fig. 3 and Supplemental Table T1 show that all strains of {\it unc-54}
mutants move with frequencies much lower than wild type. These mutants
have affected muscle myosin in the body wall, which presumably increases
the time constant controlling contraction-relaxation cycle of 
myosin-actin crossbridge. Thus, increase in muscle time constant leads
to a pronounced decrease in the undulatory frequency both for forward
and backward motion.

\vspace{0.5cm}

\noindent {\it Dependence of undulatory frequency on muscle
calcium signaling.}
We studied locomotion of {\it C. elegans} mutants with defective calcium
signaling in body wall muscle cells.
Null {\it unc-68} mutants, which have defective
ryanodine receptor channels responsible for release of stored calcium
ions, display abnormal locomotion (Maryon et al, 1996; Maryon et al, 1998). 
We analyzed the motion of loss-of-function {\it unc-68(r1158)} 
worms with reduced levels of calcium in muscles, and found that their
frequencies of undulations are smaller than that of the wild-type worms 
(Suppl. Table T1 and  Fig. S1), in agreement with Maryon 
et al. (1996, 1998).

We also analyzed locomotion of loss-of-function {\it egl-19(n582)}
animals carrying a transgene expressing EGL-19-YFP in neurons
({\it unc-119::egl-19::yfp}). This strain has a defective voltage-gated
Ca$^{+2}$ channel in body wall muscles but has its function restored in 
neurons. These worms have less muscle calcium influx (Jospin et al, 2002) 
and lower frequency of undulations than wild type (Suppl. Table T1 and 
Fig. S1), similarly to the {\it unc-68} worms.

These data suggest that decreasing calcium signaling in muscle cells
leads to decrease in the frequency of body oscillations. 
Moreover, because {\it unc-68} is expressed almost exclusively in muscles 
(Maryon et al, 1998), and {\it egl-19(n582); unc-119::egl-19::yfp}
has a defect also only in muscles, these results provide an additional
support for our hypothesis about the existence of a feedback 
coupling between body wall muscles and the neural circuits.

\vspace{1.5cm}

\noindent {\bf Theoretical results}

\vspace{0.3cm}

The major theoretical findings detailed below are: (i) the dynamic properties
of the global circuit practically do not depend on CPG design in the head 
if stretch receptor feedback is sufficiently strong and slow,
(ii) the neuromuscular wave is generated due to a nonlocal stretch receptor 
feedback coupling in the body wall, and its wavelength is almost frequency 
and stretch receptor coupling (if strong enough) independent, 
(iii) the frequency of oscillations depends in a biphasic
manner on both inhibitory and excitatory synaptic coupling, 
(iv) the frequency increases monotonically with muscle gap-junction coupling
strength, and (v) the frequency decreases with increasing time constants 
of the circuit elements. We also predict that frequency should
depend non-monotonically on muscle calcium signaling.
We made no attempt to fit quantitatively the experimental data, since
many physiological parameters in {\it C. elegans} are unknown. In this
sense, our approach is qualitative. However, we adjusted model parameters,
especially $\tau_{s}$, to obtain frequencies of the same order of magnitude 
as in experiments.

\vspace{0.3cm}

\noindent {\it Oscillation generation.}
We assume that primary oscillations necessary for a regular movement
(i.e., spontaneous and not caused by an outside mechanical stimulation) 
are generated in the head. The arguments, based on the anterior-to-posterior 
gradient of the flex (Fig. 3), supporting the head as a likely 
primary CPG are presented in the Discussion. 
We consider two extreme design versions of the head CPG.
In model A, the stretch receptor feedback plays a minor modulatory
role in affecting neurally generated rhythm (neural component dominates).
On the other hand, in model B, the stretch receptor feedback plays a critical
role in rhythm generation (reflex component dominates).
In model A oscillations are generated in the head interneuron circuit, i.e., 
in the loop X $\rightarrow$ Y $\rightarrow$ Z $\rightarrow$ X, provided 
nonlinearities in the interactions between these interneurons are 
sufficiently strong (corresponding sigmoidal functions 
${\bf H_{\alpha\beta}}$ have sharp thresholds for activation). 
The emerging oscillatory patterns are presented in Fig. 7A,C.
One of the roles of the stretch receptor feedback in model A is to
slow down the emerging rhythm as compared to a purely neural 
oscillator by introducing temporal delays.
(This is because the characteristic membrane time constant of 
{\it C. elegans} neurons should be of the order of 100 msec, as it is in 
morphologically similar {\it Ascaris} neurons (Davis and Stretton, 1989) 
and much larger mammalian neurons. As a result, a purely neural rhythm would 
give  $\sim 10$ Hz body undulations, which is 20-30 times larger than that 
observed experimentally; Suppl. Table T1.)
In model B, the primary oscillations in the head are generated because of 
the presence of the stretch receptor feedback and its interaction with
neurons, i.e. in the loop 
X $\rightarrow$ Y $\rightarrow$ E $\rightarrow$ M $\rightarrow$ S 
$\rightarrow$ X. Oscillations in this model are generated 
even when the interneuron circuit (X, Y, Z) is intrinsically 
non-oscillatory, i.e., when interactions
between interneurons are only weakly nonlinear or linear. 
In this case, however, the prerequisite for the appearance of oscillations
is initial asymmetry in activities between ventral and dorsal sides,
which is necessary for engaging the stretch receptor feedback.
Such activity asymmetry can be facilitated by nonequal number of
dorsal and ventral motor neurons present in {\it C. elegans}.
The emerging oscillatory patterns in model B are presented in Fig. 7B,D.

In both models of CPG, the crucial element for oscillation 
generation is a net non-linear negative feedback loop destabilizing 
the steady state and stabilizing a limit cycle (periodic state).
The frequency of the emerging oscillations depends on neuronal and 
non-neuronal parameters characterizing the oscillatory circuit.
The constant input $c_{1}$ (see Eq. (1) in Materials and Methods) 
coming to the head interneurons from other head neurons or/and sensory 
input can abolish oscillations if it is either too inhibitory or too 
excitatory. Thus, depending on the activity level of the head neural 
networks, a worm can decide when to start or to stop moving.
Our theoretical analysis (below) shows that models A and B give very
similar results for all model parameters, except for the coupling 
strength and time constant of the stretch receptors.

Oscillations generated in the head are imposed on the body wall
nervous system primarily through interconnections of the body wall muscles 
with the head muscles, and then via body wall stretch receptors and
synapses. However, the electric coupling between muscles and synaptic
coupling between neurons cannot be too weak, since then 
oscillations along the body wall are unsustainable and their amplitude 
(waveform) decays with a distance from the head (Fig. 7A,B, lower panels). 
In our model the stability for the oscillations is provided
by the interneurons AVB and PVC. Since these neurons receive a massive
oscillatory input with different phases in the head (from X, Y, Z), 
their resulting activities are only weakly- or non-oscillatory, because 
of a partial or complete cancellation of the input oscillatory phases.
Nearly constant input from AVB and PVC to excitatory motor neurons 
overcomes their inhibition from the DVA neuron and stabilizes 
undulatory oscillations, with ventral and dorsal sides in anti-phase 
(see also Discussion). This input in combination with sufficiently strong
synapses and gap-junctions ensure that the oscillatory waveform does not
decay with a distance from the head. Consequently, the distribution of 
frequencies along the body is homogeneous despite the fact that rhythm 
is generated in the head. This homogeneity is consistent with the experimental 
results for wild-type and all non-neuronal mutants (Figs. 3, 4 and 7; 
Suppl. Table T1). For the distribution of frequency of neuronal mutants
with weak synapses see the Discussion.

\vspace{0.5cm}

\noindent {\it Wave generation and wavelength constancy.}
Undulatory movement also requires a neuromuscular wave 
(Karbowski et al, 2006). Our model for forward locomotion can generate 
such a wave due to a non-local stretch receptor activity transduced to 
the body wall somas of excitatory motor neurons via their long unidirectional
dendrites (Fig. 2). With a posterior stretch receptor coupling the wave 
propagates posteriorly (corresponding to forward motion), while with 
an anterior stretch receptor coupling the wave propagates anteriorly 
(corresponding to backward motion) (Fig. 8A,B). Thus, the directionality
of the wave (and thus motion direction) depends only on the directionality
of the long dendrites. Unidirectional stretch receptor feedback along
the body wall is critical for locomotion, since when it is absent the 
wave does not propagate. In this case oscillations are also abolished 
in model B, although in model A they are still present (due to an 
intrinsically oscillatory interneuron loop 
X $\rightarrow$ Y $\rightarrow$ Z $\rightarrow$ X). 
The correlation between directions of the wave and a hypothetical stretch 
receptor coupling is consistent with the neuroanatomical details of the 
corresponding body wall excitatory motor neurons of type A and B, controlling
backward and forward motion, respectively. These neurons have long 
undifferentiated dendrites directed in opposite directions 
(White et al, 1986). The phase differences between neighboring 
muscle activities along the body wall are to a large extent stable over 
time and essentially frequency independent (Fig. 8C). Since, in general,
a phase lag is inversely proportional to wavelength, this implies that 
the wavelength is also frequency independent. The intermuscle phase lag 
depends, however, on the stretch receptor coupling, although very weakly 
if coupling is sufficiently strong (Fig. 8D).

\vspace{0.5cm}

\noindent {\it Biphasic dependence of frequency on synaptic coupling.}
Chemical coupling via synapses is one of the critical parameters in the 
circuit, since it can potentially affect numerous kinematic characteristics. 
Our model makes predictions about the influence of inhibitory and excitatory 
transmission on frequency. In Fig. 9A,B, we plot the dependence 
of the frequency on the inhibitory synaptic strength at the neuromuscular 
junction. This dependence exhibits a non-monotonic (biphasic) relationship.
For weak inhibitory strength, the frequency
is small and it increases with increasing inhibition up to some maximal 
value, and then it decreases.

With respect to excitation, our model shows that frequency is 
a biphasic function of the global excitatory synaptic strength 
(Fig. 9C,D), similar to the above case of inhibition. 
When the excitation level is either above or below some critical values, 
the oscillations become unstable and movement disappears. For the 
consistency of these results with the experimental data 
see the section ``Comparison of the theory with the experimental data''.

The origin of the biphasic relationships between the frequency and
inhibitory and excitatory synaptic strengths lies in the shapes of temporal
muscle voltage characteristics (Fig. 9E,F). 
For weak inhibition (or strong excitation) muscle voltage 
increases sharply but its relaxation phase is prolonged
as compared to the case of intermediate levels of inhibition
(or excitation). In contrast, for strong inhibition (or weak excitation)
the muscle voltage increases slowly but its decay phase is much faster. 
In both regimes, corresponding longer decay or rise phases of 
the muscle voltage cause the frequency drop in comparison to the 
intermediate regime (Fig. 9E,F).

\vspace{0.5cm}

\noindent {\it Frequency vs. gap-junction coupling.}
Electrical coupling via gap junctions plays also a significant role in
coordination of the circuit components. Models A and B both predict that 
frequency is a monotonically increasing function of the muscle gap junction 
conductance strength (Fig. 10A,B). Moreover, when this gap junction coupling 
is too weak, the oscillatory signal coming from the head fails to propagate 
down the body, i.e., the neuromuscular wave is attenuated, causing defective 
locomotion. On the other hand, when this coupling is too strong the wave
disappears because the whole system is synchronized. Thus, for a robust wave
the gap-junction coupling should be in an intermediate range.

\vspace{0.5cm}

\noindent {\it Frequency vs. muscle calcium signaling.}
In our model, the amount of calcium influx in muscles is controlled by 
the coupling $w_{mm}$  (Eqs. 6 and 12). In Fig. 10C,D, we display the 
biphasic or triphasic dependence of frequency on this amplitude. 
At some very high level of calcium signaling oscillations become unstable 
and the system converges to a steady-state, in which undulatory motion 
is abolished. The curves in Fig. 10C,D are qualitatively distinct 
from the biphasic curves for global excitation and inhibition presented 
in Fig. 9, since the latter exhibit pronounced local maxima.

\vspace{0.5cm}

\noindent {\it Frequency vs. stretch receptor coupling.}
One of the roles of the stretch receptor feedback coupling is to modulate 
the undulatory rhythm. We find that models A and B differ only for a weak
stretch receptor coupling. However, if the coupling is sufficiently strong,
both models show that frequency increases slightly with increasing the 
coupling (Fig. 11).

\vspace{0.5cm}

\noindent {\it Frequency vs. time constants.}
We also tested how the frequency of oscillations depends on various
time constants associated with different dynamic variables. In general,
an increase in any of the time constants causes decrease in frequency, 
with a variable sensitivity among different time constants.
As an example, in Fig. 12 we show the dependence of frequency on
the muscle membrane time constant $\tau_{m}$ (Fig. 12A), and on the
stretch receptor time constant $\tau_{s}$  (Fig. 12B). 
Models A and B differ only for small values of time constants.

\vspace{1.5cm}

\noindent {\bf Comparison of the theory with the
experimental data}

\vspace{0.3cm}

Below we make a qualitative comparison of several of our theoretical
results to either our data or to the data already published.

\begin{itemize}

\item {\it Invariance of neuromuscular wavelength.} 
This experimental result (Table 2; see Karbowski et al. (2006) for a larger
data set) is consistent with our theory, which shows that intermuscle 
phase lag is practically frequency independent (Fig. 8C), and is only 
weakly stretch receptor coupling dependent if this coupling is 
sufficiently strong (Fig. 8D). The latter result can explain
wavelength constancy across different nematode species 
(Karbowski et al, 2006).

\item {\it Biphasic dependence of frequency on synaptic 
coupling.}
For inhibitory coupling this experimental result (Fig. 5) is consistent
with our theory (Fig. 9A,B). Moreover, the theoretical curve suggests that 
wild-type worms may have GABA coupling parameters close to the optimal 
values. 

\vspace{0.2cm}

In the case of excitatory coupling, data in Fig. 6A for acetylcholine 
signaling are not in conflict with the theoretical biphasic relationship
(Fig. 9C,D), although they are not conclusive.
However, the support for the biphasic relation between frequency
and acetylcholine signaling comes from combining the experimental results 
in Fig. 6B, containing {\it slo-1} worms with transgenes, with our 
previously published data (Karbowski et al, 2006) on hyperactive 
{\it goa-1(n1134)}, {\it goa-1(sy192)}, and {\it egl-30(tg26)} mutants 
(Segalat et al, 1995; Mendel et al, 1995). Both of these groups 
have increased acetylcholine synaptic signaling above wild-type level 
but their corresponding undulatory frequencies show the opposite behavior. 
In the case of {\it slo-1} worms with transgenes, the frequency slightly 
decreases as compared to animals with the wild-type levels of 
neurotransmission (Fig. 6B), while in the case of hyperactive mutants  
the frequency increases (Karbowski et al, 2006). This behavior clearly 
suggests a non-monotonic, possibly biphasic dependence, in agreement with 
our theoretical results (Fig. 9C,D).

\item {\it  Frequency increases (decreases) monotonically 
with increasing (decreasing) muscle gap-junction coupling.} 
This theoretical result (Fig. 10A,B) agrees with the recent 
experimental results on {\it unc-9(fc16)} mutants (Liu et al, 2006). 
These mutants have body wall muscle gap junction conductance 
severely reduced (but neurotransmission at neuromuscular junction 
unaffected) and velocity much lower than wild type (Liu et al, 2006). 
As velocity and frequency are linearly related 
(see Karbowski et al, 2006), this implies that in {\it unc-9(fc16)} worms
the frequency is also strongly reduced, as our theory predicts (Fig. 10A,B).

\item {\it  Frequency decreases monotonically 
with increasing model time constants.} 
This theoretical result (Fig. 12) is consistent with the data on 
{\it unc-54} mutants. These mutants have significantly increased time constant 
characterizing the duration of the cross-bridge cycle formation between 
myosin and actin (Moerman and Fire, 1997) and their undulatory
frequency is much smaller than wild type (Fig. 3A,B; Suppl. Table T1).
In our model, Eqs. (1-13), 
this slowing down of the contraction-relaxation cycle corresponds to 
an increase of the muscle membrane time constant $\tau_{m}$.

\end{itemize}

\newpage
%\vspace{1.5cm}

\noindent {\bf\Large Discussion}

\vspace{0.5cm}

In this paper, we use a combined approach involving genetic perturbations, 
quantitative behavioral tracking, and mathematical modeling to
uncover the mechanisms of {\it C. elegans} locomotion. Our circuit model
is detailed enough to be biologically realistic and yet not excessively
complex to obscure the basic mechanisms. We hope this model provides a
link between molecular/cellular and systems level of understanding.

\vspace{0.3cm}

\noindent {\bf Discussion of the main results}

\vspace{0.3cm}

We propose the existence of a mechanosensory feedback in {\it C. elegans} 
that strongly affects locomotion. Feedback coupling is implied by our 
locomotory data on non-neuronal mutants (Fig. 4; Suppl. Table T1), in 
which neural function is not affected directly and yet they exhibit abnormal 
locomotion as compared to wild type. In particular, these non-neuronal
mutants have significantly modified frequency and flex of undulations, 
which are indirectly linked to the activation levels of the motor neurons. 
The conclusion about the existence of the mechanosensory feedback is also 
compatible with previous experimental data suggesting that altered
mechano-sensory channels can profoundly affect locomotory pattern. 
For example, {\it C. elegans} {\it unc-8} mutants, which 
have defective putative mechanosensory channels, display abnormal
locomotion (Tavernarakis et al, 1997). Also recent results on the 
mechanosensory TRPN channel TRP-4 show that it is involved in nematode 
bending control (Li et al, 2006).
Experimental data indicate that some neurons serve 
as mechanosensors with specialized sensing channels, mostly in the head 
and tail (Goodman and Schwarz, 2003; Suzuki et al, 2003; Wicks et al, 1996).
In the theoretical model of the locomotory circuit, we implemented the
mechanosensory feedback as a stretch receptor coupling in the head and
in the body wall. Its role, however, is different in both of these circuits.
In the head, the stretch feedback  participates in the control or generation
of oscillations, while in the body wall, in addition, it generates 
the neuromuscular wave by breaking a spatial longitudinal symmetry of 
the system through unidirectional coupling between body wall components. 
Possible neural candidates 
for the stretch receptors are: a pair of SAA neurons in the head, 
and VB and DB motor neurons along the body wall, because of their long
undifferentiated processes (White et al, 1986).

The remarkable experimental finding (Fig. 3 and Suppl. Table T1) 
is that the opposite motions - forward and backward - exhibit spatial 
gradients of flex and in the case of neuronal mutants also weak 
frequency gradients directed in the same way, i.e., from the head to the 
tail. This finding might suggest that primary oscillations are generated 
in the same location for both types of motion. 
Where could this location be? Because in most cases
signals propagating through a biological tissue get attenuated as
a result of energy dissipation caused by tissue viscosity, the most
likely candidate for a primary pacemaker (CPG) in {\it C. elegans} is the
head. Below, we argue that other potential candidates for the primary 
CPG, i.e. either tail or body wall or some combination, are less likely.

If tail served as a primary oscillator, then according to Fig. 3, 
this would require amplification of the signal
as it travels from the tail to the head, which in turn, would require
large amounts of mechanical energy generated in the body wall to
compensate for energy loss. Small number and low linear density of the body
wall motor neurons (37 forward motion motor neurons (VB, DB, VD, DD) 
cover $\sim$ 80 $\%$ of the body length as compared to $\sim$ 200 
neurons in the head; see White et al, 1986), 
and the decaying trend of synaptic density toward the tail 
(Fig. 7B in Chen et al, 2006) are probably not sufficient to excessively 
activate body wall muscles and achieve signal amplification.
Another possibility is that both head and tail are two separate primary
pacemakers with two distinct spontaneous frequencies. However, because
head and tail have numerous recurrent synaptic connections, these two
potential oscillators would likely adjust their frequencies to a single one,
as happens in a system of coupled oscillators (e.g. see Strogatz, 1994).
Some neuronal mutants show that head and tail oscillate differently, which
is a strong evidence against such a scenario. It is important to stress
that our hypothesis about the primary oscillator in the head does not
exclude the tail from being a secondary oscillator. This
potential oscillator could be activated only by an outside source,
either by head neurons or by a mechanical stimulation (e.g. tail tapping). 
In fact, since the tail has the second largest density of neurons in 
{\it C. elegans} (Chen et al, 2006), it is potentially capable of a slight 
signal amplification, as is evident from flex data for some mutants near 
their tails (signal amplification is possible because a typical
muscle cell in the tail should get more excitation than a typical muscle 
cell in the body wall due to a higher neural density in the tail). 
Finally, there also exists a theoretical possibility that the body wall 
motor neurons serve as a primary CPG. This hypothesis is, however, not 
compatible with our experimental results on posterior gradients of body 
undulations, i.e., one would expect more uniform spatial distributions. 
Other arguments against this possibility are: the non-zero flex of GABA 
null mutants (Table T1 in the Suppl. Information) and older data 
on interneuron ablation (Chalfie et al, 1985).
GABA null mutants have functionally disabled ventral-dorsal 
cross-inhibition at the neuromuscular junction along the body wall
(such cross-inhibition is necessary for body undulations). 
Despite this deficiency, they have non-zero flexes, which is a clear 
indication that anti-phase undulations must be imposed from another part of 
the body. By the same token, ablation of the two forward motion interneurons
AVB and PVC have been shown to virtually abolish (or strongly reduce) body 
wall undulations and the forward motion but still keep intact the head 
undulations (Chalfie et al, 1985).
This experiment suggests that forward motion body wall oscillations are not 
sustainable without the presence of connections to the head (which again 
is in support of our hypothesis that the head serves as a primary CPG). 
We want to stress, however, that the body wall motor neurons might  
serve as a secondary short-term pacemaker when stimulated from outside, 
similar to the tail.

The most interesting consequences of our theoretical model are 
the biphasic nature of the dependence of frequency on the global
strength of synaptic coupling (Fig. 9) and on muscle calcium signaling 
(Fig. 10C,D), as well as the constancy of neuromuscular wavelength
(Fig. 8C,D). The biphasic relationships, being a direct consequence 
of the non-linearities present in the system, are in agreement with 
the experimental data on GABA signaling (Fig. 5), and acetylcholine
signaling (Fig. 6) if combined with a previously published kinematic 
results on hyperactive acetylcholine mutants (Karbowski et al. 2006).
These results may suggest that wild-type worms have synaptic 
couplings close to their optimal values.
The stability of the neuromuscular wave is determined by the ability of 
the locomotory circuit to maintain the precise phase relationships 
between body components. Our model shows that unidirectional stretch 
receptor coupling involving long dendrites of the body wall motor neurons
(Fig. 2) provides a mechanism for such a coherency, since 
the phase differences between neighboring muscles and consequently 
the wavelength are stable and practically frequency and stretch
receptor coupling independent (Fig. 8C). The invariance of the wavelength
agrees well with the experimental data in Table 2 and in Karbowski et al
(2006).

Within our theoretical model, we consider two extreme forms of CPG in the 
head circuit. In model A, the primary oscillations are generated on the level 
of head interneurons and then only modulated by stretch receptor feedback, 
and ultimately by the whole locomotory network, which makes oscillations 
more robust. In model B, oscillations in 
the head emerge as a result of collective loop interactions between 
head neurons, muscles and stretch receptors, without prior oscillatory 
activity of neurons. In this scenario, the stretch receptor feedback is 
a critial element in creating oscillations, unlike in model A. 
These two mechanisms can be realized in the same circuit with the same 
topology, only with different parameters characterizing head interneurons
interactions, and worms can potentially use either of these mechanisms 
by modulating their respective parameters (change in only few of them 
is needed to transit between models A and B). 
The presence of oscillations can be controlled by sensory input that can 
change the level of excitation coming to head interneurons, 
which is controlled in our model by the parameter $c_{1}$. 
In general, both excessive excitation (large positive $c_{1}$) and excessive 
inhibition (large negative $c_{1}$) abolish the 
oscillations and consequently the motion. From a mathematical perspective, 
oscillations in the circuit are the result of non-linear feedback loops, 
which cause temporal delays in the system and make the steady state 
unstable. This mechanism is quite general and robust (insensitive on the
precise values of parameters) and appears in many biological
and non-biological (e.g. engineering) systems (Strogatz, 1994). 
It is important to stress that both of the above 
potential forms of CPG are indistinguishable on a phenotypic level, 
if stretch receptor coupling is sufficiently strong and slow (Figs. 9-12).
In this regime, both models yield qualitatively similar predictions 
regarding dynamic properties of the locomotory circuit under different 
perturbations, and both are consistent with the experimental data. 
This fact suggests that global locomotory output is to a large extent a result 
of the collective dynamics of the whole circuit, not just its local elements, 
and it does not depend on a specific mechanism of oscillation generation
in most of the parameter space. In our model, the head interneuron circuitry
is treated in a simplified manner given the complexity of the actual
circuitry. This simplified approach, however, captures an important feature
of interneuron connectivity pattern, namely small-scale feedback loops.
The model head interneurons X, Y, Z form a minimal circuit (loop) in
{\it C. elegans} that can generate robust oscillations.
In general, our results suggest that more experimental attention should 
be paid to the head circuit in the context of locomotion.

\vspace{0.3cm}

\noindent {\bf Discussion of additional aspects of this study}

\vspace{0.3cm}

Our theoretical study shows the existence of 
several ``windows of parameters'' (usually sufficiently wide) for the 
presence of undamped oscillations along worm's body, which is
required for a sustained undulatory motion. 
Beyond those critical ranges of parameters oscillations decay just 
after a few cycles, with the most pronounced decay for posterior parts of 
the circuit. For example, the global levels of excitation and inhibition 
in the network can be neither too small nor too large for the stable 
oscillations (Fig. 7). The weak frequency gradients we observe 
experimentally for neuronal mutants with compromised synaptic transmission 
(Figs. 3, 5 and 6; Suppl. Tables T1, T2) may be directly related to the 
spatial decay of the waveform along the body, which we notice in
simulations (Fig. 7A,B lower panels). In particular, our simulations 
indicate that for mutants with weak synaptic coupling the more
posterior a given muscle is the more damped are its oscillations.
It seems that our automated experimental setup has problems with detecting
low amplitude oscillations (low signal-to-noise ratio) in more posterior
parts, which manifests itself in weak frequency gradients.

Similarly, muscle calcium signaling cannot be too large to maintain undulatory 
locomotion. Experiments by Maryon et al. (1996, 1998), in which wild-type 
worms were exposed to ryanodine, are consistent with the last prediction. 
Worms treated with ryanodine, which produces an excessive amount of calcium 
in muscle cells, are either very sluggish or completely paralyzed 
(Maryon et al, 1996).

In our model, the activity of AVB and
PVC has two other important functions. First, to make oscillations in 
the body wall circuit (imposed from the head via muscles and stretch 
receptors) more robust by releasing excitatory motor neurons (VB and DB) 
from inhibitory influence of DVA (when this inhibition is too strong 
or a signal from AVB and PVC too weak oscillations are not sustained). 
Second, to start the posterior neuromuscular wave by signaling to VB and 
DB motor neurons with the help of their long posterior dendrites.
This wave is a second critical factor for the forward locomotion, 
as is evident from a biomechanical model of worm's undulations 
(Karbowski et al, 2006). 
Following the Byerly and Russell hypothesis we assume that long dendrites
contain stretch receptors that provide posterior coupling between
spatially separated parts of the body wall. Without this coupling,
the wave is abolished in both models A and B.
The involvment of stretch receptors in generation
of the wave was also suggested before on purely biomechanical grounds 
(Niebur and Erdos, 1993).
It is important to stress that forward and backward motion require 
neuromuscular waves traveling in opposite directions, which are correlated
with the corresponding opposite directions of the motor neuron's dendrites 
in forward and backward circuits. However, based on the above arguments
it seems that both types of motion have CPG's located probably in the head.

In the model, spontaneous collective oscillations can be also generated 
in the body wall circuit without the involvement of the head circuit 
and/or AVB and PVC interneurons, because of the numerous feedback loops 
present in the body wall. This mechanism was theoretically considered by
Bryden and Cohen (2004). We did not, however, pursue this possibility
for the reasons outlined above. Therefore, the parameters in our model 
are chosen such that the body wall neural circuit does not oscillate in 
isolation from the head circuit, i.e., when AVB and PVC neurons are removed 
and when there is no coupling between muscles in the head and the body wall. 
In particular, we assumed that the excitatory (VB, DB neurons) and 
inhibitory (VD, DD neurons) motor neurons are intrinsically non-oscillatory.

Our theoretical results show, for both models A and B, that
frequency increases weakly with increasing the stretch receptor
coupling if it is strong enough (Fig. 11). 
This theoretical result is consistent with data for 
{\it trp-4} mutants (Li et al, 2006) and cuticle {\it sqt-1} mutants.
These animals show the opposite behavior, i.e.,  {\it trp-4} are
faster and have larger flex than wild-type (Li et al, 2006),
while {\it sqt-1} are slower and have smaller flex (Suppl. Table T1). 
Decreased (increased) flex angle implies smaller (larger) stretching, 
which in turn suggests weaker (stronger) stretch receptor feedback coupling, 
and according to our theory, lower (higher) frequency.
The case of BE109 mutants is more complex, since they exhibit higher
flex angle and lower frequencies than the wild-type worms (Suppl. 
Table T1). The likely explanation in this case is that despite higher flexes 
the actual stretch receptor coupling might be smaller than in wild-type.
This is possible because these mutants are noticeably shorter than
wild type (about 20$\%$), suggesting a profound structural changes 
in the cutico-skeletal system that might affect the quality of 
a mechano-sensory signaling.

{\it C. elegans} locomotion seems similar to the locomotion of other
systems performing undulations. However, there are also some important 
differences, in particular, other undulatory systems such as leech, 
lamprey, and crayfish have anatomically well defined body
segments, which serve as local CPG's (Skinner and Mulloney, 1998; 
Friesen and Cang, 2001; Marder et al, 2005). 
In contrast, {\it C. elegans} lacks body segments (White el al, 1986)
and its body wall seems to be rather passive in comparison to the head,
which likely serves as CPG.

We note that our circuit model allows a direct investigation of the effects
of single- or multi-neuron ablations (e.g. motor neurons) on locomotion 
characteristics. By comparing kinematic parameters of {\it C. elegans} 
with ablations
to circuit model predictions one can in principle gain information
on the strength of some synaptic connections and refine the model.

Some of our extrachromosomal transgene constructs involving {\it slo-1}
mutants might not precisely restore the wild-type functions. However,
because our analysis of the experimental data is to a large extent
qualitative, the possible under- or over-shooting to the wild-type 
level are not critical to our conclusions.

\vspace{0.3cm}

\noindent {\bf Conclusions}

\vspace{0.3cm}

Our major conclusions are: (i) based on the anterior-to-posterior 
gradients of the flex for almost all strains, and partly based on the 
corresponding weak gradients of the frequency for neuronal mutants, 
the most likely location of CPG for both forward and backward locomotion 
is the head, (ii) stretch receptor feedback associated with long 
unidirectional dendrites in the body wall plays
a critical role in neuromuscular wave generation and conservation of
the wavelength, (iii) there exist optimal values of synaptic couplings
(both inhibitory and excitatory) for which frequency of undulations
is maximal, and wild-type worms might be close to that optimal regime,
(iv) the precise mechanism of oscillation generation within CPG, i.e.,
whether neural activity or reflex activity dominate, is irrelevant and
indistinguishable on a phenotypic level, except for weak and fast stretch 
receptor coupling, (v) frequency increases monotonically with muscle 
gap-junction strength, and (vi) frequency decreases monotonically with
increasing time constants of the circuit elements.
Our circuit model can also make predictions: (i) frequency should depend
in a non-monotonic manner on muscle calcium signaling, and (ii) long dendrites
of the excitatory motor neurons in the body wall are critical for generating
neuromuscular wave and thus perturbing them should perturb the wavelength.   
Experiments addressing these problems would either provide additional
support for our model or cast doubt on some assumptions, thus helping
in refining the model.

\vspace{1.2cm}

\noindent{\bf Acknowledgments}

We thank A. Davies and S. McIntire for {\it slo-1} strains, and 
J. Kramer for advice on cuticle mutants. The work was supported by the 
Sloan-Swartz Fellowship (J.K.), by NIH Fellowship (NS043037 to G.S.), 
by USPHS grant R01-DA018341 (P.W.S.), and by the Howard Hughes Medical 
Institute, with which P.W.S. is an Investigator.

\vspace{1.4cm}

{\bf References}

\vspace{0.3cm}

\noindent Akay T, Haehn S, Schmitz J, Buschges A (2004). Signals
from load sensors underline interjoint coordination during stepping
movements of the stick insect leg. {\it J. Neurophysiol.} 92: 42-51.

\noindent Bargmann CI (1998). Neurobiology of the 
{\it Caenorhabditis elegans} genome. {\it Science} 282: 2028-2033.

\noindent Brenner S (1974). The genetics of {\it Caenorhabditis elegans}.
{\it Genetics} 77: 71-94.

\noindent Bryden J, Cohen N (2004). A simulation model of the locomotion
controlers for the nematode {\it Caenorhabditis} elegans. In: Schaal S 
et al editors. From Animals to Animats 8: Proc. Eight Intern. Conf. on 
Simulation of Adaptive Behavior. Cambridge: MIT Press, pp 183-192.

\noindent Chalfie M, White J (1988). {\it The Nervous System.} 
In: Wood WB, editor. The Nematode {\it Caenorhabditis elegans}.
Cold Spring Harbor: Cold Spring Harbor Laboratory Press, 
pp. 337-391.

\noindent Chalfie M et al (1985). The neural circuit for touch 
sensitivity in {\it Caenorhabditis elegans.} {\it J. Neurosci.} 5: 956-964.

\noindent Chen BL, Hall DH, Chklovskii DB (2006). Wiring optimization can
relate neuronal structure and function. {\it Proc. Natl. Acad. Sci. USA}
103: 4723-4728.

\noindent Cronin CJ et al. (2005). An automated system for
measuring parameters of nematode sinusoidal movement. {\it BMC Genet.}
6: 5.

\noindent Davis RE, Stretton AOW (1989). Signaling properties
of {\it Ascaris} motorneurons: graded active responses, graded synaptic
transmission and tonic transmitter release. {\it J. Neurosci.} 9: 
415-425.

\noindent Davies AG et al (2003). 	
A central role of the BK potassium channel in behavioral responses to 
ethanol in {\it C. elegans}. {\it Cell} 115: 655-666.

\noindent Delcomyn F (1980). Neural basis of rhythmic behavior in animals.
{\it Science} 210: 492-498.

\noindent de Bono M, Maricq AV (2005). Neuronal substrates of
complex behaviors in {\it C. elegans}. {\it Annu. Rev. Neurosci.}
28: 451-501.

\noindent Francis MM, Mellem JE, Maricq AV (2003). Bridging the gap
between genes and behavior: recent advances in the electrophysiological
analysis of neural function in {\it Caenorhabditis elegans}.
{\it Trends Neurosci.} 26: 90-99.

\noindent Friesen WO, Cang J (2001). Sensory and central mechanisms
control intersegmental coordination. {\it Curr. Opin. Neurobio.} 11: 678-683.

\noindent Gengyo-Ando K et al. (1993). The {\it C. elegans}
{\it unc-18} gene encodes a protein expressed in motor neurons.
{\it Neuron} 11: 703-711 (1993).

\noindent Goodman MB, Hall DH, Avery L, Lockery SR (1998).
Active currents regulate sensitivity and dynamic range in {\it C. elegans}
neurons. {\it Neuron} 20: 763-772.

\noindent Goodman MB, Schwarz EM (2003). Transducing touch in
{\it Caenorhabditis elegans.} {\it Annu. Rev. Physiol.}  65:
429-452.

\noindent Gray JM, Hill JJ, Bargmann CI (2005). A circuit for navigation
in {\it Caenorhabditis} elegans. {\it Proc. Natl. Acad. Sci. USA} 102: 
3184-3191.

\noindent Grillner S (1975). Locomotion in vertebrates: central mechanisms 
and reflex interaction. {\it Physiol. Rev.} 55: 247-303.

\noindent Hobert O (2003). Behavioral plasticity in {\it C. elegans}:
Paradigms, circuits, genes. {\it J. Neurobiol.} 54: 203-223.

\noindent Jin Y et al (1999). The {\it Caenorhabditis elegans} gene
{\it unc-25} encodes glutamic acid decarboxylase and is required for
synaptic transmission but not synaptic development. {\it J. Neurosci.}
19: 539-548.

\noindent Jospin M, Jacquemond V, Mariol MC, Segalat L,
Allard B (2002). The L-type voltage-dependent Ca$^{2+}$ channel EGL-19
controls body wall muscle function in {\it Caenorhabditis elegans}. 
{\it J. Cell Biol.} 159: 337-347.

\noindent Karbowski J, Cronin CJ, Seah A, Mendel JE, Cleary D,
Sternberg PW (2006). Conservation rules, their breakdown, and optimality
in {\it Caenorhabditis} sinusoidal locomotion. {\it J. Theor. Biol.}
242: 652-669.

\noindent Lee RYN, Lobel L, Hengartner M, Horvitz HR, and
Avery L (1997). Mutations in the $\alpha$1 subunit of an L-type 
voltage-activated Ca$^{2+}$ channel cause myotonia in 
{\it Caenorhabditis elegans}. {\it EMBO J.} 16: 6066-6076.

\noindent Li W, Feng Z, Sternberg PW, Xu XZS (2006). A {\it C. elegans}
stretch receptor neuron revealed by a mechanosensitive TRP channel
homologue. {\it Nature} 440: 684-687.

\noindent Liu Q, Chen B, Gaier E, Joshi J, Wang Z-W (2006).
Low conductance gap junctions mediate specific electrical coupling
in body-wall muscle cells of {\it Caenorhabdidtis elegans.} 
{\it J. Biol. Chem.} 281: 7881-7889.

\noindent Maduro M, Pilgrim D (1995). Identification and cloning 
of unc-119, a gene expressed in the {\it Caenorhabditis 
elegans} nervous system. {\it Genetics} 141: 977-988.

\noindent Marder E, Calabrese RL (1996). Principles of rhythmic motor
pattern generation. {\it Physiol. Rev.} 76: 687-717.

\noindent Marder E, Bucher D, Schulz DJ, Taylor AL (2005).
Invertebrate central pattern generation moves along. {\it Curr Biology}
15: R685-R699.

\noindent Maryon EB, Coronado R, Anderson P (1996). {\it unc-68}
encodes a ryanodine receptor involved in regulating {\it C. elegans}
body-wall muscle contraction. {\it J. Cell Biol.} 134: 885-893.

\noindent Maryon EB, Saari B, Anderson P (1998). Muscle-specific
functions of ryanodine receptor channels in {\it Caenorhabditis elegans}.
{\it J. Cell Sci.} 111: 2885-2895.

\noindent McIntire SL, Jorgensen E, Kaplan J, Horvitz HR (1993).
The GABAergic nervous system of {\it Caenorhabditis elegans}. {\it Nature} 
364: 337-341.

\noindent Mendel JE et al (1995). Participation of the protein Go
in multiple aspects of behavior in C. elegans.  Science 267: 
1652-1655.

\noindent Moerman DG, Fire A (1997). {\it Muscle: Structure, Function,
and Development. } In: Riddle DL et al. editors. C. elegans II. 
Cold Spring Harbor: Cold Cold Spring Harbor Laboratory Press, 
pp. 417-470.

\noindent Niebur E, Erdos P (1991). Theory of the locomotion of
nematodes. {\it Biophys. J.} 60: 1132-1146.

\noindent Nusbaum MP, Beenhakker MP (2002).  A small-systems
approach to motor pattern generation. {\it Nature} 417: 343-350.

\noindent Okuda T, Haga T, Kanai Y, Endou H, Ishihara T, Katsura I. 
(2000). Identification and characterization of the high-affinity choline 
transporter. {\it Nature Neurosci.} 3: 120-125.

\noindent Schuske K, Beg AA, Jorgensen EM (2004). The GABA
nervous system in {\it C. elegans}. {\it Trends Neurosci.} 27: 407-414.

\noindent Segalat L, Elkes DA, Kaplan JM (1995). Modulation
of serotonin-controlled behaviors by Go in Caenorhabditis elegans.
{\it Science}  267: 1648-1651.

\noindent Skinner FK, Mulloney B (1998). Intersegmental coordination
in invertebrates and vertebrates. {\it Curr. Opin. Neurobiol.} 8: 725-732.

\noindent Strogatz SH (1994). {\it Nonlinear Dynamics and Chaos}.
Westview Press.

\noindent Suzuki H et al (2003). In vivo imaging of {\it C. elegans}
mechanosensory neurons demonstrates a specific role for the MEC-4
channel in the process of gentle touch sensation. {\it Neuron}  39:
1005-1017.

\noindent Tavernarakis N, Shreffler W, Wang S, Driscoll M 
(1997). unc-8, a DEG/ENaC family member, encodes a subunit of a candidate
mechanically gated channel that modulates {\it C. elegans} locomotion.
{\it Neuron} 18: 107-119.

\noindent Wang ZW, Saifee O, Nonet ML, Salkoff L (2001). Slo-1
potassium channels control quantal content of neurotransmitter release
at the {\it C. elegans} neuromuscular junction. {\it Neuron} 32: 867-881.

\noindent Weimer R.M., et al. (2003). Defects in synaptic vesicle
docking in {\it unc-18} mutants. {\it Nature Neurosci.} 6: 1023-1030.

\noindent White JG, Southgate E, Thomson JN, Brenner S (1986).
The structure of the nervous system of the nematode 
{\it Caenorhabditis elegans.} {\it Phil. Trans. R. Soc. Lond.} B 314: 
1-340.

\noindent Wicks SR, Roehrig CJ, Rankin CH (1996). A dynamic network
simulation of the nematode tap withdrawal circuits: Predictions concerning
synaptic function using behavioral criteria. {\it J. Neurosci.} 
16: 4017-4031.

%\newpage

\begin{table}
\begin{center}
\caption{Parameters used in simulations of Eqs. (1)-(13).}
\begin{tabular}{|l l|l l|l l|}
\hline
\hline
Parameter &     &   Parameter &     &  Parameter &       \\
\hline
\hline

$\tau_{x}$ & 130 ms & $\tau_{y}$ & 100 ms & $\tau_{z}$ & 150 ms \\ 
$\tau_{e}$ & 150 ms & $\tau_{i}$ & 120 ms & $\tau_{m}$ & 200 ms \\ 
$\tau_{s}$ & 350 ms & $\tau_{avb}$ & 165 ms & $\tau_{pvc}$ & 150 ms \\

\hline
 
$c_{1}$ &  1.0 & $c_{2}$ &  $-1.0$ &    &  \\

\hline

${\tilde w}_{me}$ & $q_{ex}$  &  ${\tilde w}_{mi}$ & $0.5q_{in}$  
&  $w_{xz}$ & 0.6  \\
$w_{zx}$  & $q_{ex}$  &  $w_{ey}$ & $0.5q_{ex}$  & $w_{xi}$ 
& $0.3{\tilde w}_{mi}$ \\
$w_{yx}$ & $0.5q_{ex}$  &  $w_{zy}$ & $0.5q_{ex}$  &  $w_{es}$ & 0.4 \\
$w_{xs}$ & $5.0w_{es}$  &  $w_{ie}$ & $1.5q_{ex}$  & $w_{me}$ & $0.5q_{ex}$ \\
$w_{mi}$ & $0.3q_{in}$  &  $w_{sm}$ & 0.5  &  $w_{mm}$ & 0.1 \\
$w_{e,pvc}$ & $0.2q_{ex}$  &  $w_{avb,pvc}$ & $0.2q_{ex}$  &  
$w_{avb,x}$ & $0.4q_{ex}$  \\
$w_{pvc,x}$ & $0.2q_{ex}$  &  &   &  & \\ 

\hline

$g_{avb}$ & 0.10  &  $g_{m}$ &  0.10  &  $g_{e}$ & 0.05  \\
$g_{i}$ & 0.05  &  $g_{avb,e}$ & 0.10  &    &  \\

\hline

$\theta_{me}$ & 0.55  &  $\theta_{mi}$ & 0.40  &  $\theta_{xz}$ & 0.25  \\
$\theta_{zx}$ & 0.35  &  $\theta_{ey}$ & 0.30  &  $\theta_{xi}$ & 0.40  \\
$\theta_{yx}$ & 0.30  &  $\theta_{zy}$ & 0.35  &  $ \theta_{es}$ & 0.20 \\
$\theta_{xs}$ & 0.20  &  $\theta_{ie}$ & 0.55  &  $\theta_{sm}$ & 0.00 \\  
$\theta_{mm}$ & 0.55  & $\theta_{e,pvc}$ & 0.45 & $\theta_{avb,pvc}$ & 0.45 \\ 
$\theta_{avb,x}$ & 0.18  &  $\theta_{pvc,x}$ & 0.20  &  &  \\

\hline

$\eta_{me}$ & 0.80  &  $\eta_{mi}$ & 1.20  & $\eta_{xz}$ & 0.02(A),0.20(B)\\
$\eta_{zx}$ & 0.05(A),0.65(B) & $\eta_{ey}$ & 0.05(A),0.80(B) 
& $\eta_{xi}$ & 1.20  \\
$\eta_{yx}$ & 0.05(A),0.80(B) & $\eta_{zy}$ & 0.05(A),0.65(B) 
&  $\eta_{es}$ & 0.03 \\
$\eta_{xs}$ & 0.03  &  $\eta_{ie}$ & 0.80  &  $\eta_{sm}$ & 0.05 \\  
$\eta_{mm}$ & 2.40  & $\eta_{e,pvc}$ & 1.00 & $\eta_{avb,pvc}$ & 1.00 \\ 
$\eta_{avb,x}$ & 0.60  &  $\eta_{pvc,x}$ & 0.60  &  &  \\

\hline
\end{tabular}
\end{center}
Inhibitory $w_{xz}$ is assumed to be non-GABA-ergic. Labels (A) and (B)
refer to models A and B.
\end{table}

%\newpage

\begin{table}
\begin{center}
\caption{Body length normalized wavelength during forward and backward 
locomotion for adult wild-type (WT) {\it C. elegans} and mutants.}
\begin{tabular}{|l l l|}
%\multicolumn{2}{l}
%{ Table 2.}\\
%\multicolumn{3}{l} 
\hline 
\hline

 Genotype   &   Normalized wavelength   & \\
           &    Forward motion  &  Backward motion \\     

\hline\hline

{\it C. elegans} WT (N=62) &    0.61$\pm$0.3   &  0.64$\pm$0.07  \\

               &    & \\

 Neuronal Mutants:  &   &   \\

 $\;$  {\it unc-25(e156)} (N=4) &    0.68$\pm$0.01  &  0.69$\pm$0.04    \\

 $\;$  {\it unc-46(e177)} (N=4) &    0.71$\pm$0.01  &    0.71$\pm$0.02  \\

 $\;$  {\it unc-18(e81); cho-1::unc-18::yfp} (N=14) &  0.70$\pm$0.03  
 &  0.71$\pm$0.07   \\

 $\;$  {\it slo-1(js118)} (N=18)   &   0.62$\pm$0.02  &   0.64$\pm$0.02   \\

               &    &  \\

 Non-neuronal Mutants: &   &      \\

 $\;$  {\it egl-19(n582); unc-119::egl-19::yfp}  &    & \\
 $\;\;\;$  Line A (N=16)    &   0.64$\pm$0.03  &   0.67$\pm$0.13     \\

 $\;\;\;$  Line B (N=16)    &   0.63$\pm$0.03  &   0.66$\pm$0.07     \\

 $\;$  {\it unc-68(r1158)}  (N=16) &  0.62$\pm$0.03  &  0.62$\pm$0.06  \\

\hline 

\end{tabular}
\end{center}
\end{table}

\newpage

{\bf \large Figure Legends}

\vspace{0.3cm}

Fig. 1\\
The general approach taken in this paper. (A) We perturb (manipulate)
genes whose products affect some properties of different classes 
of neurons and other non-neuronal parameters. These altered properties, 
in turn, affect locomotory characteristics. The locomotory pattern may 
affect activities of neurons via mechanosensory feedback.
(B) Actual image of a worm with superimposed reference points measuring
worm's coordinates. Based on a spatio-temporal pattern of each of these
points we calculated motion characteristics. The maximal angle over
time between two solid lines (green and red on-line) determines the half 
of a local flex angle, which quantifies a deviation from a straight line of 
two neighboring reference body sections during motion. Dashed arrow (blue
on-line) indicates the direction of motion). Adapted from Cronin et al. (2005).

\vspace{0.3cm}

Fig. 2\\
Schematic diagram of the circuit for forward locomotion in the nematode
{\it C. elegans}. Excitatory synapses are denoted 
by arrows ($\rightarrow$), inhibitory synapses by bars ($\dashv$), and
electric gap junctions by dashed lines ($- -$). 
(A) The large scale view of the circuit. Head neurons excite
and inhibit head muscles (ventral $M_{hv}$ and dorsal $M_{hd}$), and
additionally receive feedback from head stretch receptors (ventral
$S_{hv}$ and dorsal $S_{hd}$). The signal from the head neurons is
transmitted to the body wall circuit through two interneurons AVB and 
PVC (with very long dendrites spanning the whole body length), and
through the gap junction coupling between head and body wall ($M_{v}$ and
$M_{d}$) muscles. Body wall excitatory motor neurons (ventral VB and
dorsal DB) use acetylcholine for synaptic signaling. These neurons
have elongated processes directed posteriorly, which, we assume, 
contain stretch receptors ($S$). Stretch receptors detect relative
differences in the activities of the corresponding ventral and dorsal
muscles. Body wall inhibitory motor neurons (ventral VD and dorsal DD)
use GABA for synaptic signaling. These neurons act as cross-inhibitors, 
since their only input is from the opposite side excitatory motor neurons. 
Excitatory motor neurons as well as inhibitory motor neurons are connected 
within each group by gap junctions (not shown).
Also, excitatory motor neurons (VB and DB) receive synaptic input from
DVA interneuron, which we assume is inhibitory. 
(B) Schematic diagram of the head CPG. The head interneurons
$X$ (corresponding to SAA neuron) and $Y$ are excitatory and, we assume, 
signal with acetylcholine, 
while $Z$ is inhibitory and, we assume, signals without GABA (GABA is
expressed only in one head interneuron RIS and its ablation does not
influence phenotype; see Schuske et al, 2004). Ventral and dorsal groups of 
these interneurons effectively inhibit each other such that oscillatory 
rhythm in the head alternates between ventral and dorsal sides. 
Head excitatory motor neurons ($E_{hv}$ and $E_{hd}$ corresponding, e.g., 
to pairs of SMB and SMD neurons) use acetylcholine, and head inhibitory 
motor neurons ($I_{hv}$ and $I_{hd}$ corresponding to a pair of RME neurons) 
use GABA for synaptic signaling. As in the body wall, stretch receptors 
in the head ($S_{hv}$ and $S_{hd}$) detect relative differences in the 
ventral and dorsal activities of head muscles ($M_{hv}$ and $M_{hd}$).
The primary oscillations can be generated either within the $X, Y, Z$ 
loop or with a stretch receptor feedback within the $X, Y, E, M, S$ loop. 
Changes in the level of global excitation (acetylcholine) and inhibition
(GABA) affect frequency of these oscillations.

%\newpage
\vspace{0.3cm}

Fig. 3\\
Experimental examples of distributions of the undulatory frequency and 
bending flex angle along the worm's body for forward and backward motion. 
(A), (B) Frequency is relatively uniform during forward and backward 
motion for wild type (solid blue line) and non-neuronal mutants 
{\it unc-68(r1158)} (dashed-dotted green line) and {\it unc-54(s74)} 
(dotted black line), but it decays weakly towards the tail for GABA 
defective {\it unc-25(e156)} mutant in forward locomotion (dashed red line). 
(C), (D) Flex angle decays markedly posteriorly for all these animals,
in some cases, with an increase towards the tail 
(notably for {\it unc-54(s74)}). 
The correlation coefficient $r$ between the flex and position from the 
head has the following values. 
For forward motion:
$r=-0.97 \; (p=0.000)$ for wild type;
$r=-0.86 \; (p=0.001)$ for {\it unc-25(e156)};
$r=-0.83 \; (p=0.002)$ for {\it unc-68(r1158)};
$r=-0.71 \; (p=0.014)$ for {\it unc-54(s74)}.
Similarly, for the backward motion:
$r=-0.89 \; (p=0.000)$ for wild type;
$r=-0.62 \; (p=0.042)$ for {\it unc-25(e156)};
$r=-0.72 \; (p=0.013)$ for {\it unc-68(r1158)};
$r= 0.41 \; (p=0.208)$ for {\it unc-54(s74)}.
In all cases but one (for {\it unc-54} the backward motion $r$
is positive but not significant) 
correlation are strong and negative, which implies that there is a
significant decreasing trend of flex with position along the body.
The line convention is the same as in (A), (B). Here and in other figures
the worm's head position corresponds to the point 1 on the
horizontal axis and its tail position to the point 11. The error bars in
all figures are population standard deviations.

\newpage
%\vspace{0.3cm}

Fig. 4\\
Locomotion characteristics of the mutants with affected non-neuronal 
functions and wild-type {\it C. elegans} (forward motion).
(A) Spatial distribution of the frequency of undulations for cuticle
mutants {\it sqt-1(sc101)} (dashed line (red on-line), $N= 13$) and BE109 
(dashed-dotted line (green on-line), $N= 9$), and wild-type 
(solid line (blue on-line), $N= 62$) 
worms. The error bars are population standard deviations. 
(B) Snapshots of worm postures: (a) wild type, (b) BE109,
(c) {\it sqt-1(sc101)}, (d) {\it unc-54(s95)}. All of the mutants
move on average slower than wild type, but their body posture and
locomotory pattern can be either similar to wild type (panel (c)) 
or significantly different (panels (b) and (d)).

\vspace{0.3cm}

Fig. 5\\
Experimental non-monotonic (biphasic) dependence of the undulatory frequency 
on synaptic GABA signaling (forward motion). (A) Loss-of-function GABA 
mutants {\it unc-25(e156)} (dashed red line, $N=4$), {\it unc-46(e177)} 
(dashed-dotted green line, $N=4$), and {\it unc-18(e81); cho-1::unc-18::yfp} 
(dotted black line, $N=14$) move with significantly lower frequencies than 
wild type (solid blue line) (non-parametric sign test for paired average 
frequencies along the body, $p < 0.05$). 
(B) Mutants {\it slo-1(js118); Pacr-2::slo-1} (dashed-dotted black
lines representing 4 independent lines A, B, C, D) 
with increased GABA signaling and presumably wild-type 
levels of acetylcholine signaling move with 
significantly lower frequencies than wild type (solid blue line) 
(sign test, $p < 0.05$).
(C) Similar to panel B: mutants {\it slo-1(js118); Pacr-2::slo-1}
(dashed-dotted black line) are also slower than two types of animals 
with presumably recovered wild-type neurotransmission in both GABA and
cholinergic neurons:
{\it slo-1(js118); Pacr-2::slo-1; Punc-25::slo-1} (solid red line, $N=21$) 
and {\it slo-1(js118); Psnb-1::slo-1} (solid green line, $N=20$) 
(sign test, $p < 0.05$).

\vspace{0.3cm}

Fig. 6\\
Experimental dependence of the undulatory frequency on synaptic 
acetylcholine signaling (forward motion).
(A) In a background with elevated GABA signaling, 
{\it slo-1(js118); Pacr-2::slo-1} worms with presumably wild-type levels of 
acetylcholine move with either slightly higher (lines A ($N=19$) 
and B ($N=18$); dashed red line) or slightly lower 
(lines C ($N=18$) and D ($N=17$); dashed green line) frequencies 
than the control {\it slo-1(js118); Pacr-2::gfp} (solid blue line, $N=19$),
which has both types of neurotransmission increased above the wild-type level. 
(B) Mutants with increased acetylcholine and presumably rescued GABA
signaling towards wild-type level, i.e., 
{\it slo-1(js118); Punc-25::slo-1} (dashed red line; $N=18$ (line A),
$N=16$ (line B)), 
{\it slo-1(js118); Punc-17::slo-1} (dashed-dotted green line; 
$N=16$ (line A), $N=18$ (line B), $N=18$ (line C)),
and {\it slo-1(js118); Pcho-1::slo-1} (dotted black line; $N=18$ (line A),
$N=18$ (line B), $N=17$ (line C)),
have slightly lower frequencies that the wild-type worms (solid blue line)
(nonparametric sign-test for paired average frequencies along the body, 
$p < 0.05$).

\vspace{0.3cm}

Fig. 7\\
Oscillations generated in the circuit.
(A) Model A. Upper panel: ventral muscle activities along the circuit 
(head $M_{hv}$ - solid line (blue on-line), $M_{v3}$ - dashed line 
(red on-line), $M_{v6}$ - dashed-dotted line (green on-line)). 
Middle panel: corresponding dorsal muscle
activities ($M_{hd}$, $M_{d3}$, $M_{d6}$) along the circuit. 
Note alternate corresponding oscillations 
in both these panels ($q_{ex}=4.0$, $q_{in}=3.0$).
Lower panel: damped oscillations along the circuit (the same line
convention as in the upper panel). For more posterior circuit elements 
oscillations become unstable only after a few cycles when either 
global excitation and inhibition are too small or muscle gap-junction
coupling is too small ($q_{ex}=2.0$, $q_{in}=1.0$). This damping effect
may explain the experimentally observed weak spatial decay of frequency 
for neuronal mutants with weak synaptic couplings (Suppl. Tables T1, T2).  
(B) The same for model B. 
(C) Oscillations in the circuit elements of model A. Upper panel:
activities of head ventral $X$ (solid line (blue on-line)) and $Z$ 
(dashed line (red on-line)) neurons. 
Middle panel: activities of two interneurons AVB (solid line (blue on-line))
and PVC (dashed line (red on-line)). Note weak high frequency oscillations 
in both interneurons in contrast to other elements in the circuit. 
Lower panel: activities of ventral body wall motor neurons (no 2)
excitatory (solid line (blue on-line)) and inhibitory 
(dashed line (red on-line)), and stretch receptors 
(dashed-dotted line (green on-line)) ($q_{ex}=3.0$, $q_{in}=2.0$).
Note that almost all elements in the circuit oscillate with the same 
collective frequency, except the two interneurons, which exhibit weak
high frequency oscillations (for some model parameters these oscillations
are practically invisible).
(D) The same for model B.

\vspace{0.3cm}

Fig. 8\\
Neuromuscular wave generated in the circuit. The wave propagates
either towards the tail (A) for posteriorly directed stretch receptor 
coupling, or towards the head (B) for anteriorly directed coupling.
The former corresponds to the forward motion circuit, whereas the
latter to the backward motion circuit. Wave does not propagate when 
a unidirectional stretch receptor feedback coupling is absent.
(C) The phase lag of neighboring muscle activities is to a large extent 
both position and frequency independent. Also the phase lag between the 
head and the tail is essentially frequency independent: 
$0.97\times(2\pi)$ for 0.70 Hz, $1.03\times(2\pi)$ 
for 0.57 Hz, and  $0.91\times(2\pi)$ for 0.46 Hz, which suggests frequency
independence of the wavelength. Solid line (blue on-line) with diamonds 
corresponds to the frequency 0.70 Hz, dashed line (red on-line) with 
circles to 0.57 Hz, and dashed-dotted line (green on-line) with crosses 
to 0.46 Hz. (D) The phase lag of intermuscle activities (between no. 3 and 2) 
decreases with increasing the stretch receptor feedback coupling, although 
weakly if the coupling is sufficiently strong for both models A 
(solid line (blue on-line)) and B (dashed line (red on-line)). 
Parameters used in (A) and (B): 
$q_{ex}=4.0$, $q_{in}=2.0$, and $g_{m}=0.4$ (model A).
Parameters used in (C) (model B only) and (D) (both models): 
$q_{ex}=3.0$, $q_{in}=2.0$.

\vspace{0.3cm}

Fig. 9\\
Theoretical dependence of the frequency of oscillations in the circuit on
the global synaptic inhibition $q_{in}$ (A, B) and global synaptic 
excitation $q_{ex}$ (C, D). Both models A and B give a qualitatively 
similar biphasic dependence of frequency on inhibition and excitation.
(A), (B) For very weak inhibition oscillations are unstable.
(C), (D) For both too weak and too strong excitation oscillations are 
unstable. (A), (B) The solid line (blue on-line) corresponds to 
$q_{ex}= 4.0$ and dashed line (red on-line) to $q_{ex}= 2.0$. 
(C), (D) The solid line (blue on-line) corresponds to $q_{in}= 5.0$ and 
dashed line (red on-line) to $q_{in}= 1.0$. 
These biphasic relationships can be explained by the changes in temporal 
characteristics of muscle voltage in response to changes in the levels
of inhibition (E) and excitation (F). Parameters used: 
(E) $q_{ex}= 4.0$ and $q_{in}= 0.5$ (upper panel), $q_{in}= 4.5$ 
(middle panel), $q_{in}= 35.0$ (lower panel) in model B; 
(F) $q_{in}= 5.0$ and $q_{ex}= 1.2$ (upper panel), $q_{in}= 3.8$ 
(middle panel), $q_{in}= 6.6$ (lower panel) in model A.

%\newpage
\vspace{0.3cm}

Fig. 10\\
Theoretical dependence of the frequency of oscillations in the circuit on
the muscle characteristics. Models A and B give qualitatively similar 
behaviors of frequency on muscle gap junction coupling $g_{m}$ (A, B) 
and muscle calcium signaling strength $w_{mm}$ (C, D). 
(A), (B) Monotonic increase of frequency with increasing gap junction 
coupling. Solid line (blue on-line) corresponds to $q_{ex}=2.0$, $q_{in}= 2.0$ 
and dashed line (red on-line) to $q_{ex}=2.0$, $q_{in}= 4.0$. 
(C), (D) Biphasic or triphasic dependence on frequency on calcium signaling.
The curve's shape depends on a relative strengths of the global excitation 
and inhibition. For very large calcium signaling levels oscillations in 
the circuit become unstable.
Solid line (blue on-line) corresponds to $q_{ex}=4.0$, $q_{in}= 4.0$ 
and dashed line (red on-line) to $q_{ex}=1.5$, $q_{in}= 7.0$.

\vspace{0.3cm}

Fig. 11\\
Theoretical dependence of the frequency of oscillations in the circuit
on the stretch receptor coupling parameter $w_{es}$. The behavior of 
model A (A) differs from model B (B) only in the weak coupling regime. 
In this regime oscillations in model B are unstable, while model A shows
unmodulated high frequency oscillations of head interneurons.
In both figures, the solid line (blue on-line) corresponds to $q_{ex}=2.0$, 
$q_{in}= 2.0$ and the dashed line (red on-line) to $q_{ex}=4.0$, $q_{in}= 1.0$.

\vspace{0.3cm}

Fig. 12\\ 
Theoretical dependence of the frequency of oscillations on the muscle 
and stretch receptor time constants.
(A) Decrease of the oscillatory frequency with increasing the muscle
membrane time constant $\tau_{m}$ in model A (solid line (blue on-line))
and model B (dashed line (red on-line)).
(B) Decrease of the oscillatory frequency with increasing the stretch
receptor time constant $\tau_{s}$ in model A (solid line (blue on-line))
and model B (dashed line (red on-line)). Note that the frequency is much 
larger in model B for small values of $\tau_{m}$ and $\tau_{s}$.
In both figures, we used $q_{ex}=3.0$ and $q_{in}= 2.0$

%\end{narrowtext}
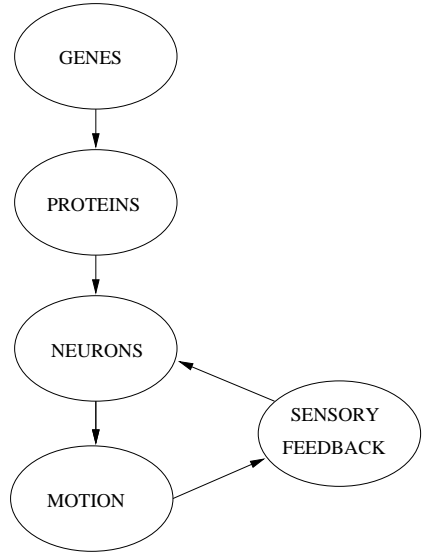
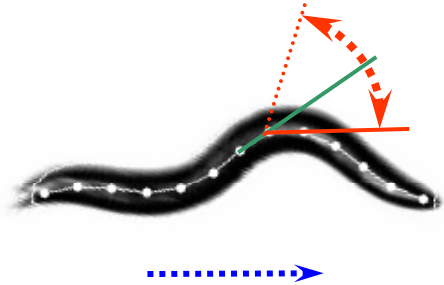
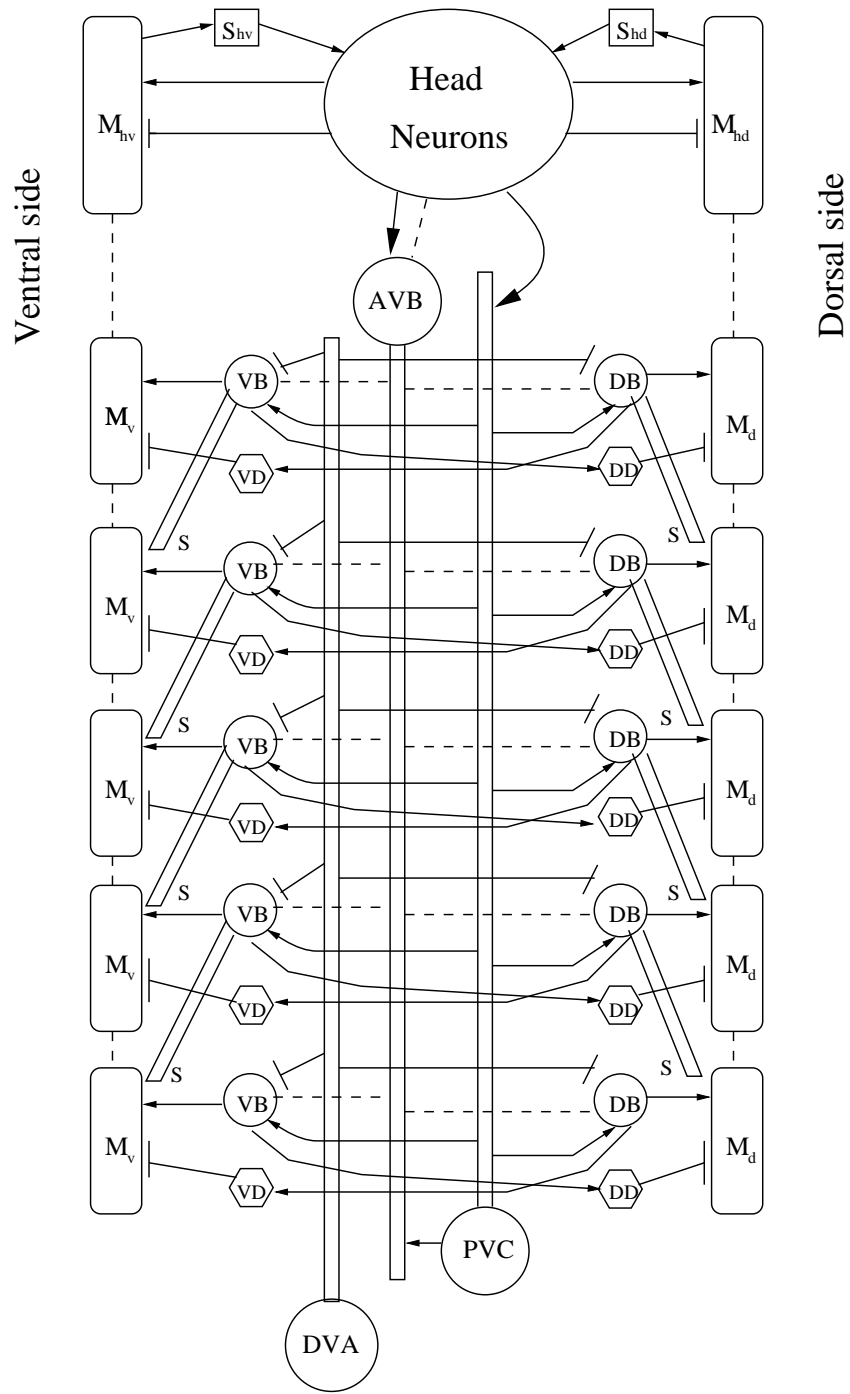
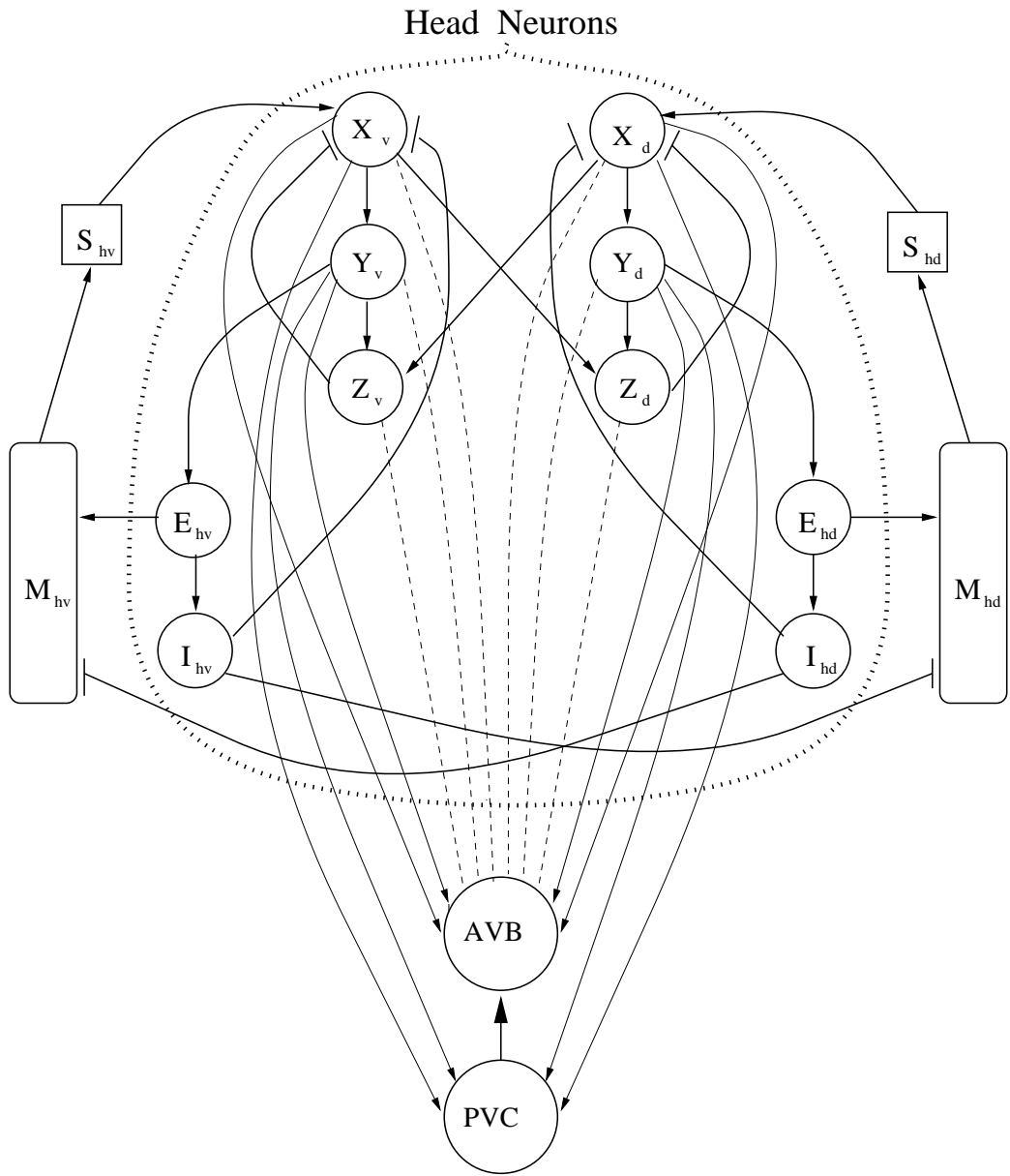
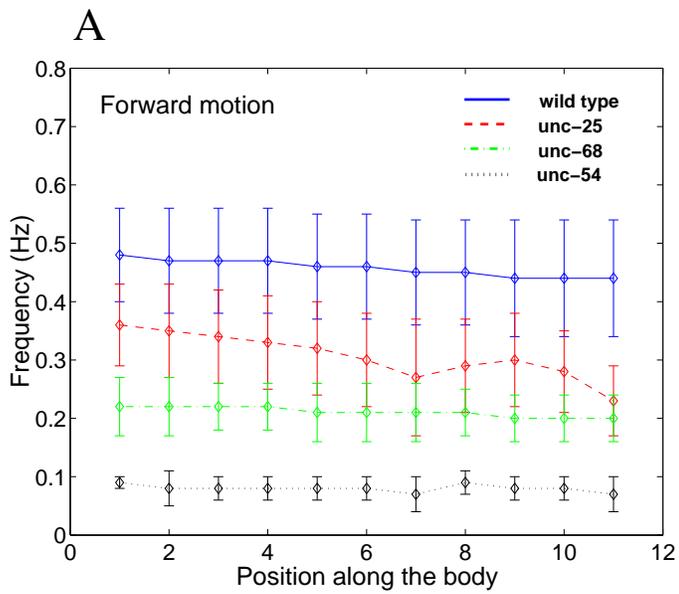
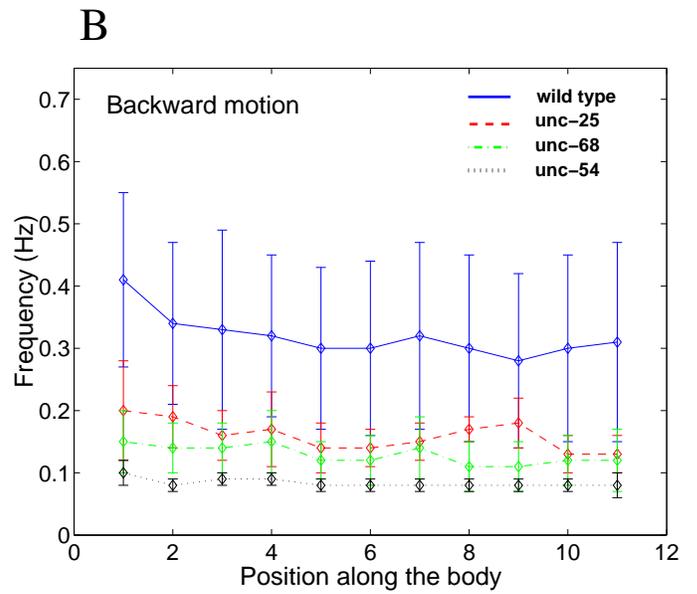
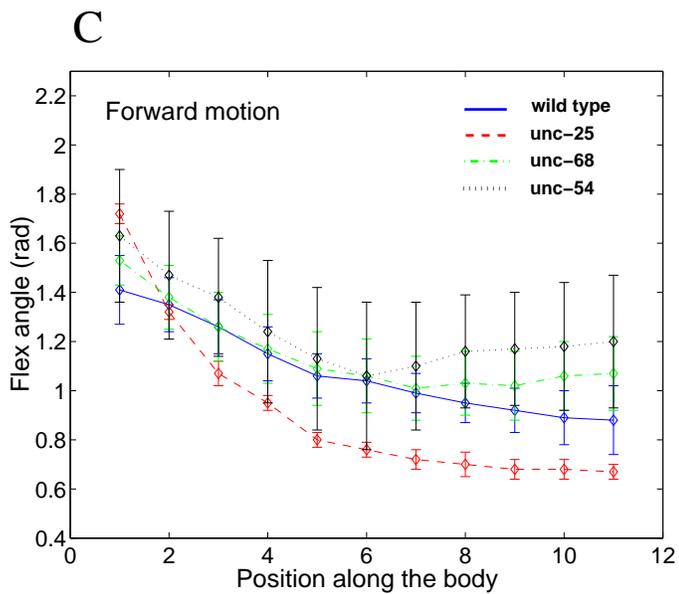
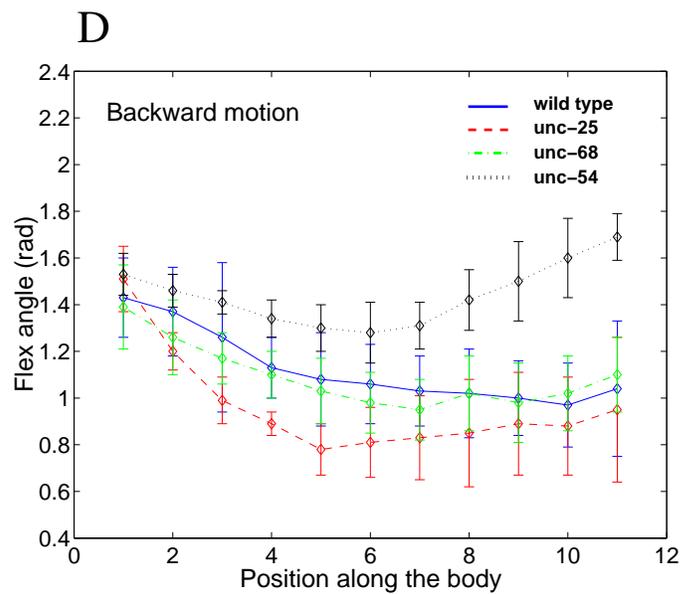
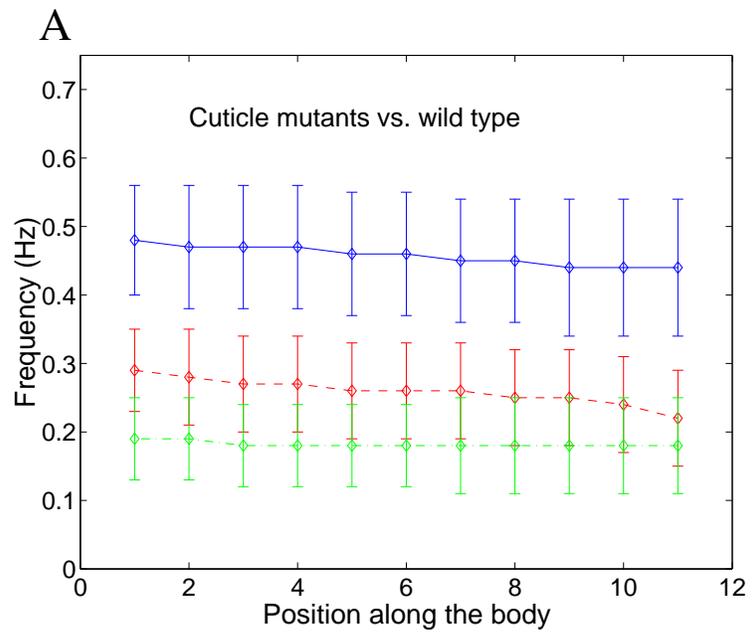
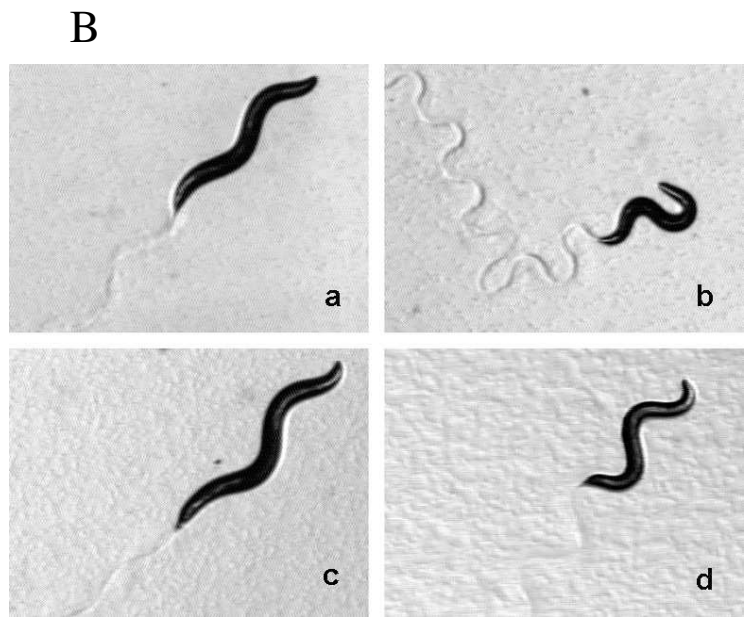
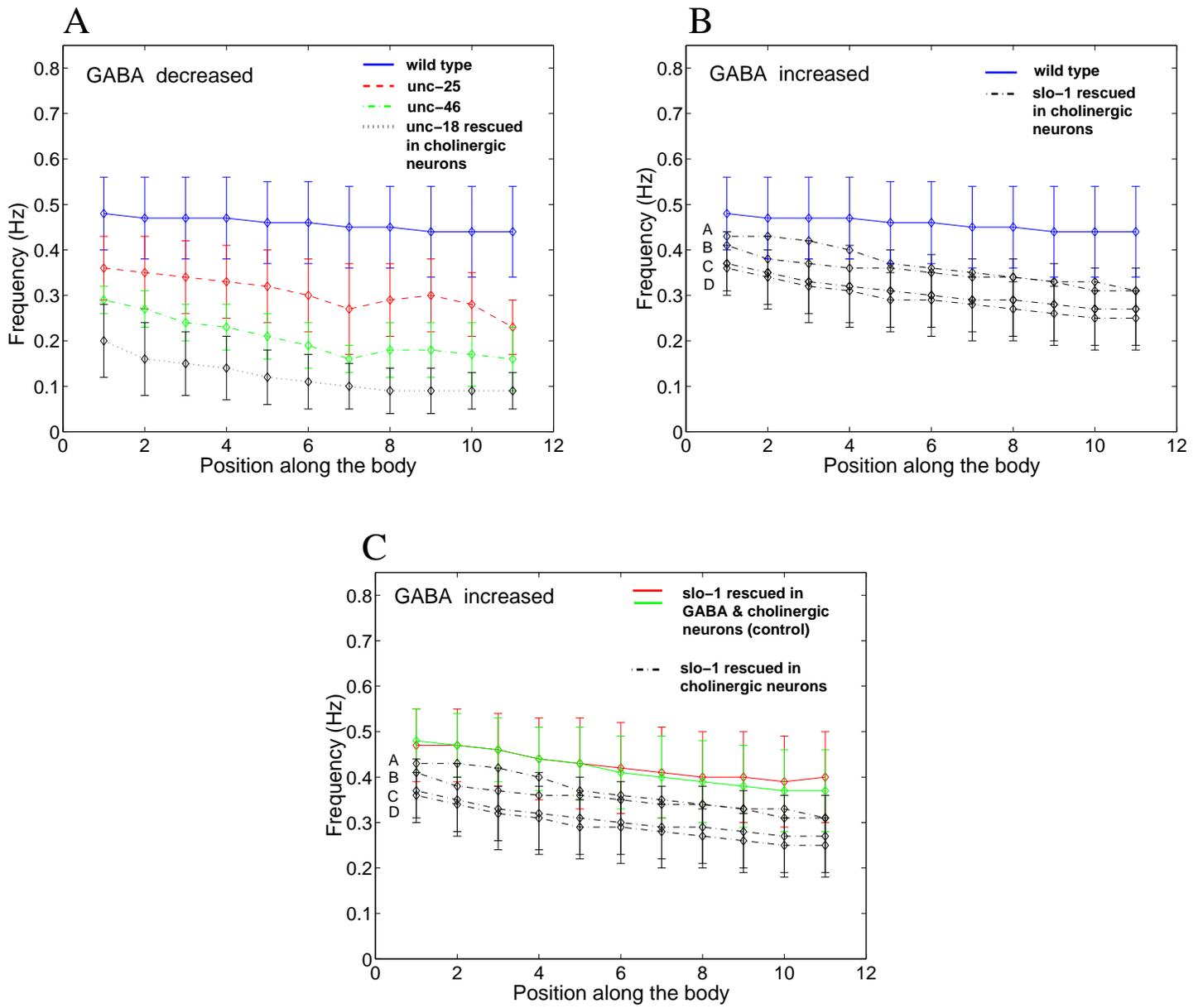
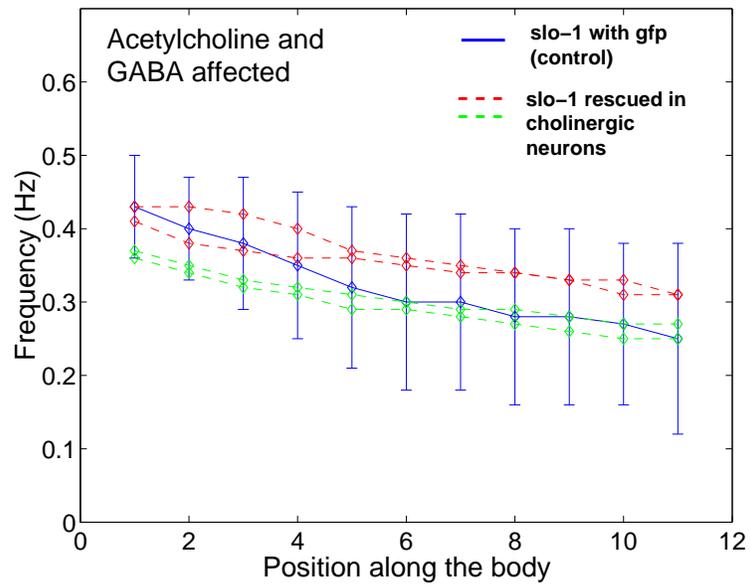
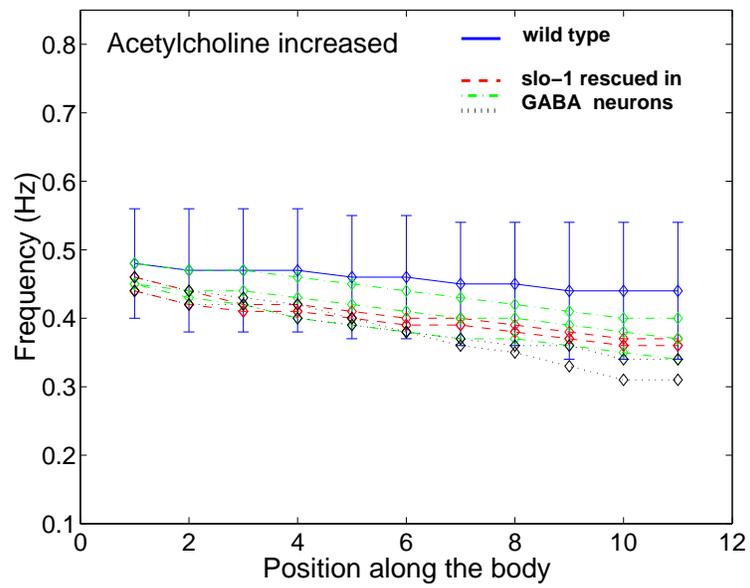
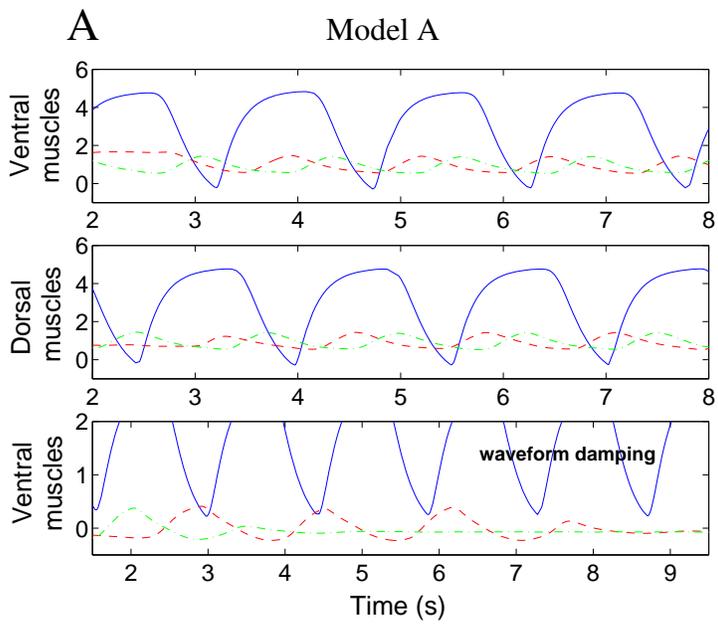
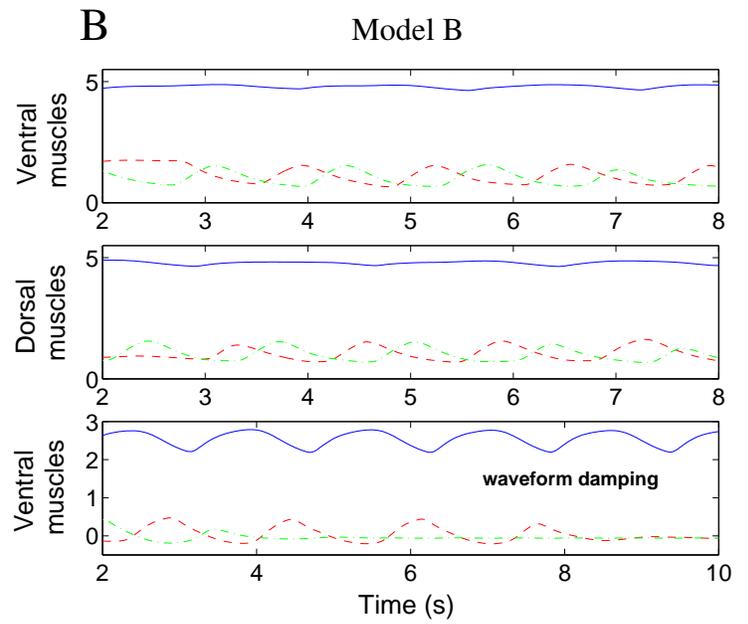
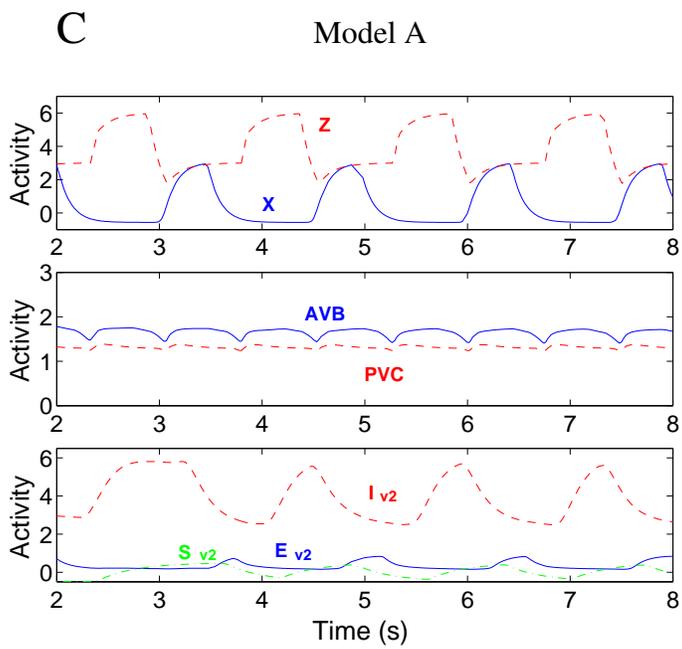
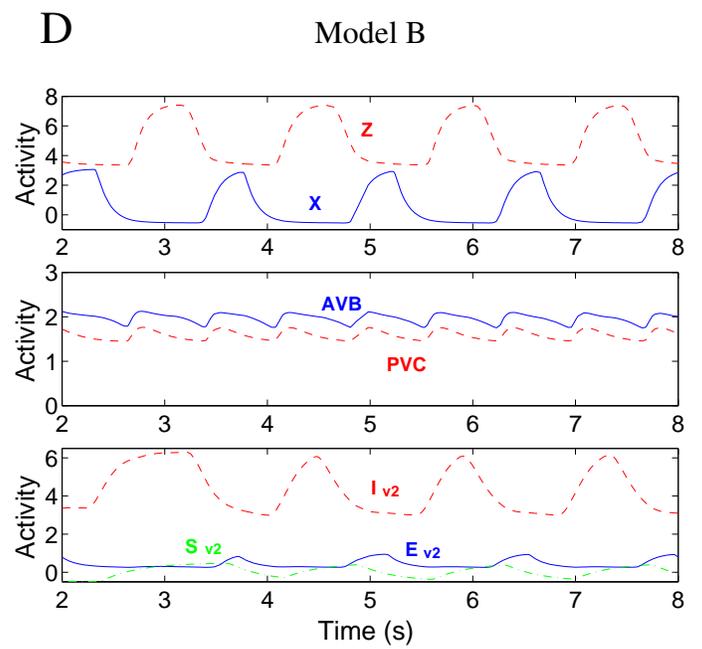
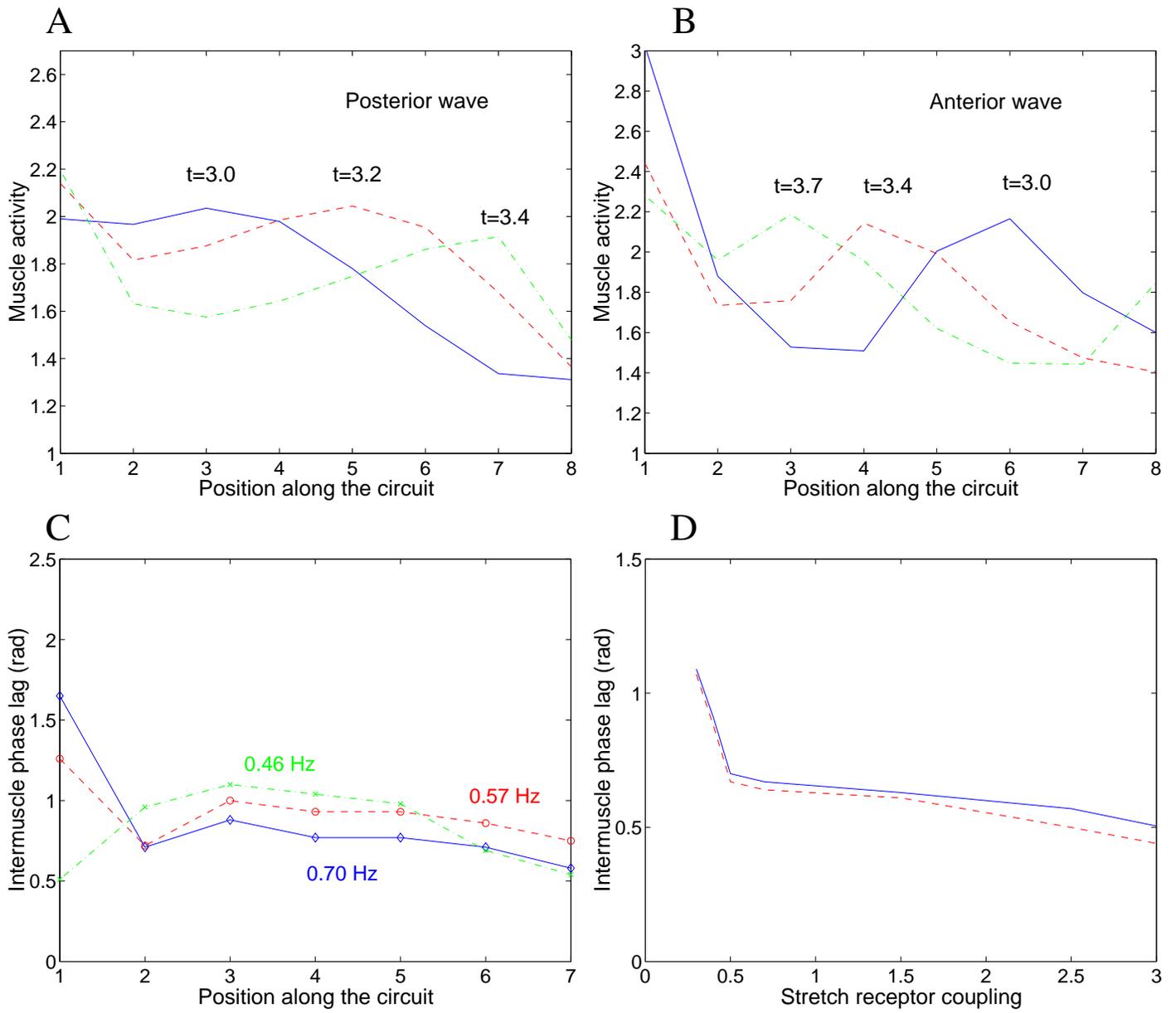
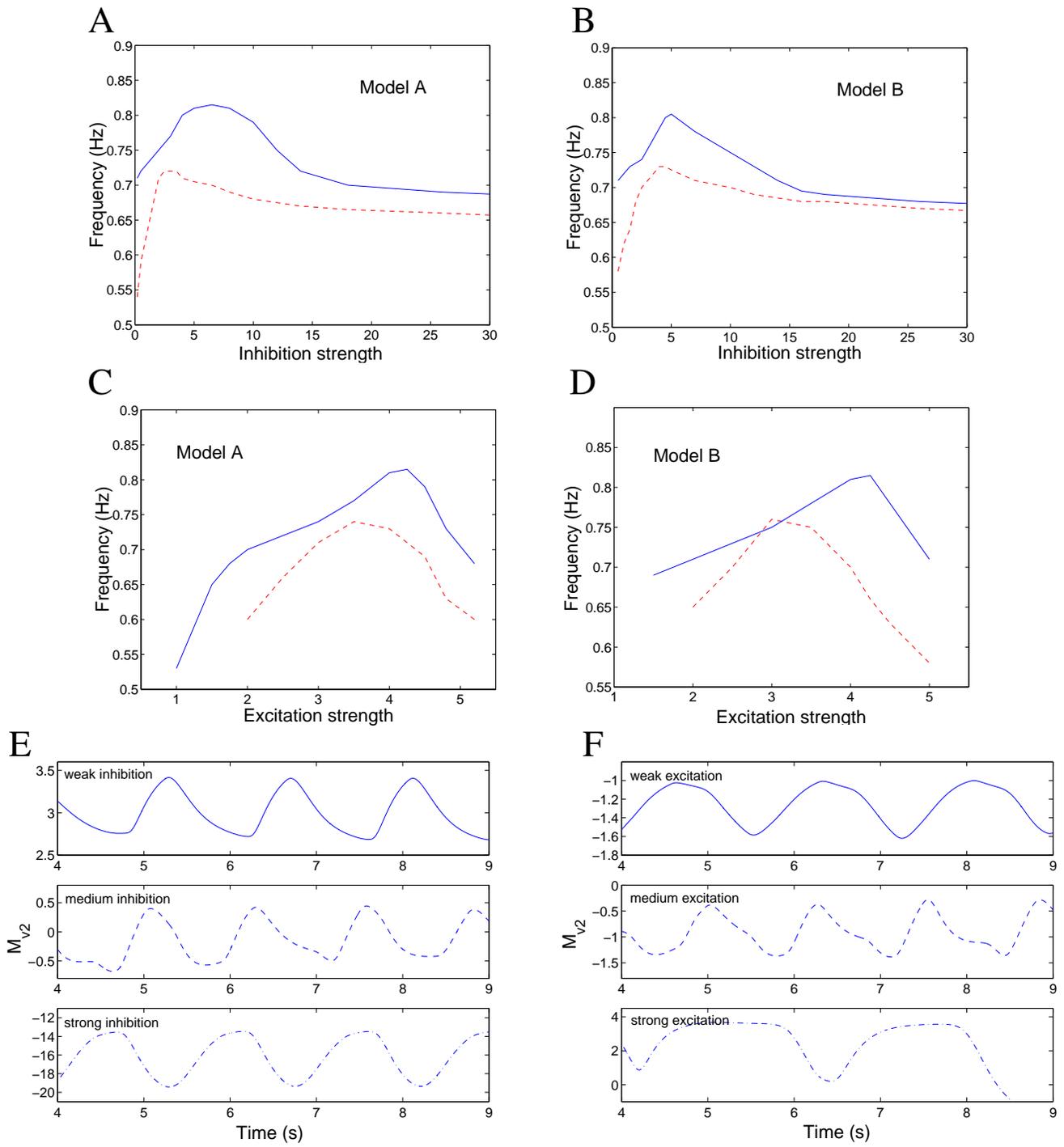
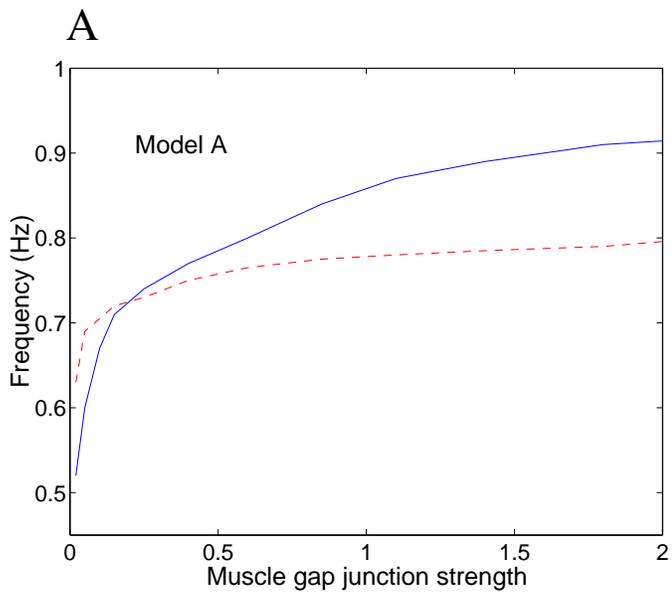
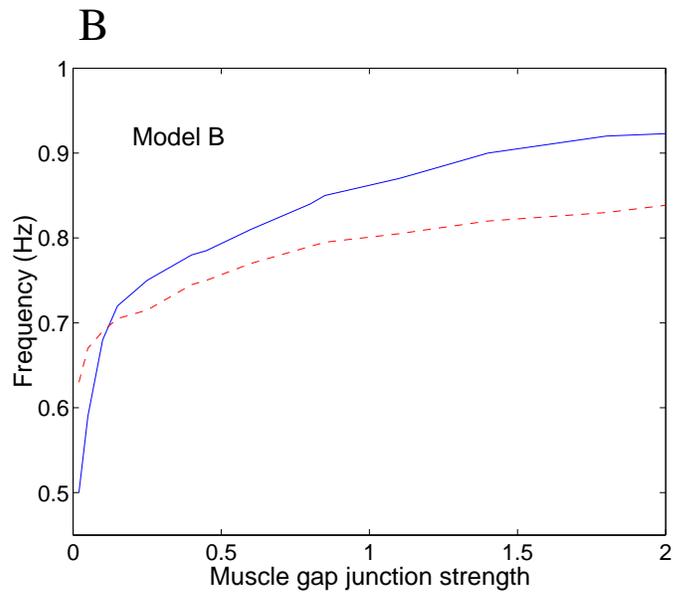
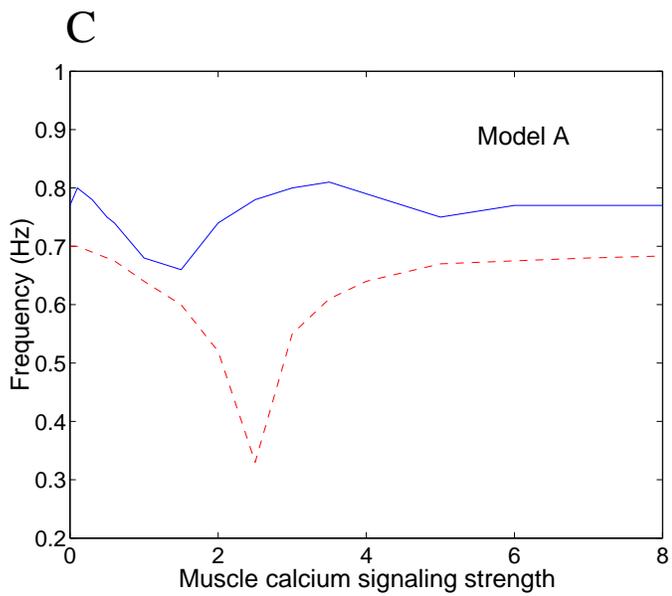
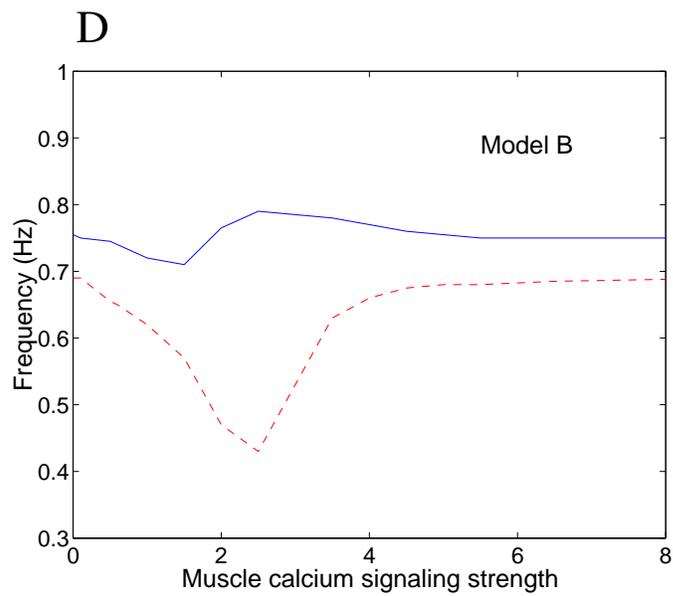
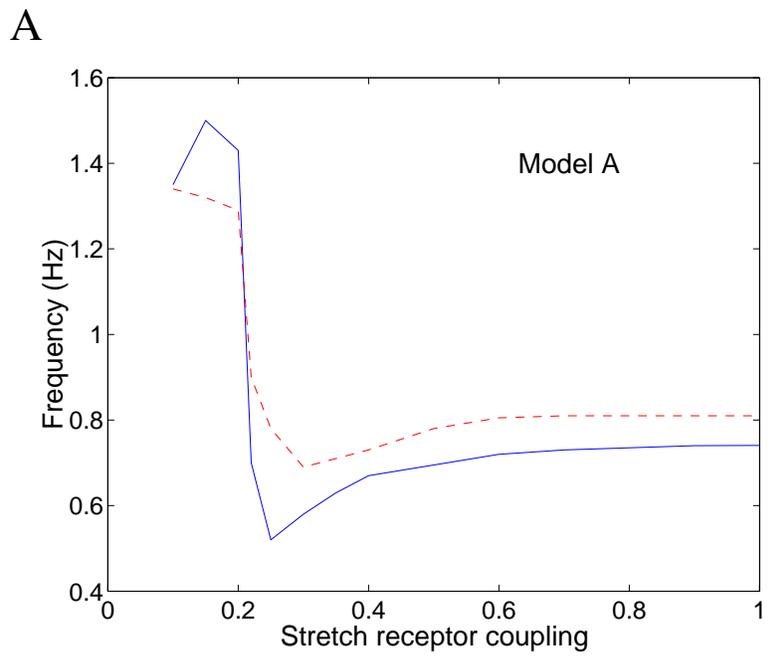
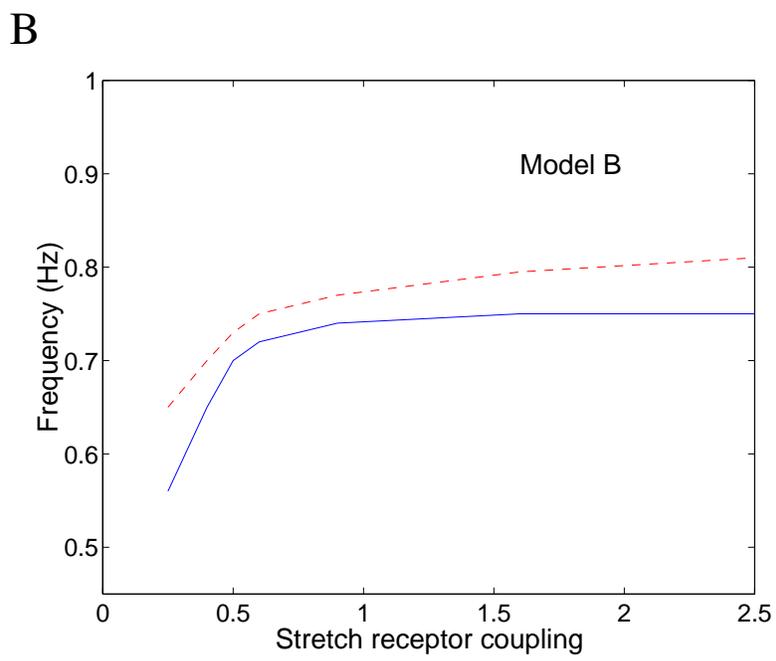
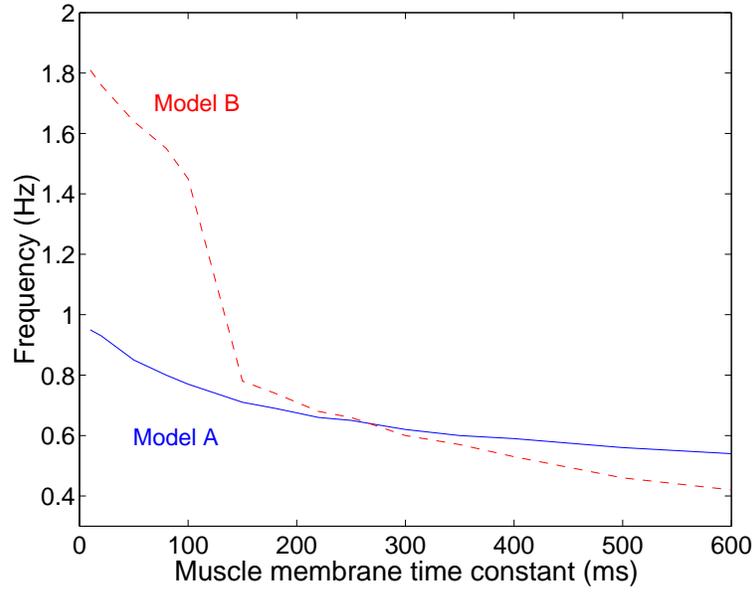
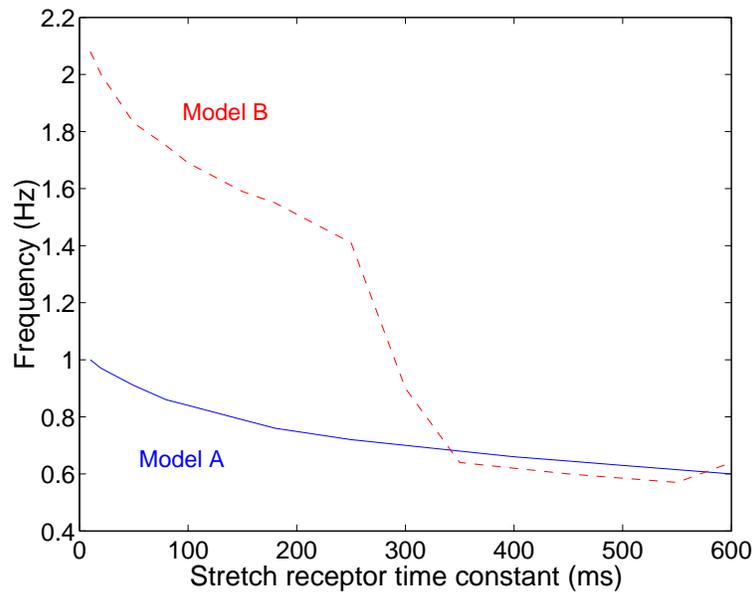
\end{document}